\documentclass[12pt,amsart]{iopart}
\pdfoutput=1
\usepackage{graphicx,psfrag,color,hyperref,amsbsy,iopams}
\providecommand{\keywords}[1]{\textbf{\textit{Keywords:}} #1}
\begin{document}
\newcommand{\gae}{\lower 2pt \hbox{$\, \buildrel {\scriptstyle >}\over {\scriptstyle
\sim}\,$}}
\newcommand{\lae}{\lower 2pt \hbox{$\, \buildrel {\scriptstyle <}\over {\scriptstyle
\sim}\,$}}

\title{Relaxation to equilibrium in models of classical
spins with long-range interactions}
\author{Debraj Das and Shamik Gupta}
\address{Department of Physics, Ramakrishna Mission
Vivekananda University, Belur Math, Howrah, 711202, India}
\ead{debraj.das@rkmvu.ac.in, shamik.gupta@rkmvu.ac.in}
\begin{abstract}
For a model long-range interacting system of classical Heisenberg spins,
we study how fluctuations, such as those arising from having a finite
system size or through interaction with the environment, affect the
dynamical process of relaxation to Boltzmann-Gibbs equilibrium. Under
deterministic spin precessional dynamics, we unveil the full range of
quasistationary behavior observed during relaxation
to equilibrium, whereby the system is trapped in nonequilibrium states
for times that diverge with the system size. The corresponding
stochastic dynamics, modeling interaction with the environment and constructed in the spirit of the stochastic
Landau-Lifshitz-Gilbert equation, however shows a fast relaxation to
equilibrium on a size-independent timescale and no signature of
quasistationarity, provided the noise is strong enough. Similar fast
relaxation is also seen in Glauber Monte Carlo dynamics of the
model, thus establishing the ubiquity of what has been reported earlier in particle
dynamics (hence distinct from the spin dynamics considered here) of
long-range interacting systems, that quasistationarity observed in
deterministic dynamics is washed away by
fluctuations induced through contact with the environment.     
\end{abstract}
\date{\today}
\keywords{Stationary states, Metastable states, Stochastic particle dynamics}
\maketitle
\tableofcontents

\section{Introduction and model of study}
\label{sec:intro}
Stochasticity is present in any statistical system constituted by a
finite number of interacting degrees of freedom, which is known to
induce fluctuations in both static and time-dependent observables of the
system, thereby affecting their statistical properties. Stochasticity
may arise due to sampling of initial conditions and due to interaction
with the external environment. It is evidently of interest
to investigate how these two sources of stochasticity interplay in
dictating the long-time state of the system, and in particular, in
predicting the values of the macroscopic observables the system
attains in the stationary state. In this work, we explore the
aforementioned issues within the ambit of a model many-body interacting system
comprising classical spins that are
interacting with one another through an inter-particle potential that is
long-ranged in nature. Namely, the interparticle potential decays rather
slowly as a
function of the separation $r$ between the particles, specifically, as
$1/r^\alpha;~0 \le \alpha \le d$, with $d$ being the embedding spatial
dimension of the
system~\cite{Campa:2009,Bouchet:2010,Campa:2014,Levin:2014,Gupta:2017}. 

Long-range interacting (LRI) systems may be
found at all length scales, from atomic to astrophysical, and are
known to exhibit a range of physical phenomena that appear
counterintuitive when viewed vis-\`{a}-vis short-range systems for which
the interaction has a finite range. A basic property that distinguishes
LRI systems from short-range ones is the violation of additivity,
whereby a macroscopic LRI system cannot be divided into independent
macroscopic subparts so that thermodynamic quantities referring to the
subparts add up to yield the corresponding values for the
composite system. While non-additivity results in such unusual features
as inequivalent equilibrium ensembles and a non-concave entropy function, more
striking is its consequence on dynamical properties, namely, that an
isolated LRI system relaxes to equilibrium over a time that diverges
with the number of degrees of freedom~\cite{Campa:2014}. Consequently, the larger the
system is, the longer is the time that it takes to attain equilibrium,
resulting in slowly-evolving nonequilibrium states being directly accessible to experimental
observations~\cite{Campa:2014}. 

Our model of study consists of $N$ globally-coupled classical Heisenberg spins
of unit length denoted by ${\mathbf S}_i=(S_{ix},S_{iy},S_{iz});~i=1,2,\ldots,N$. Expressed in terms of spherical polar angles $\theta_i \in
[0,\pi]$ and $\phi_i \in [0,2\pi)$, one has $S_{ix}=\sin \theta_i \cos
\phi_i,~S_{iy}=\sin \theta_i \sin
\phi_i,~S_{iz}=\cos \theta_i$. The Hamiltonian of the system is given by
\begin{equation}
H=-\frac{J}{2N}\sum_{i,j=1}^N {\bf S}_i \cdot {\bf
        S}_j+D\sum_{i=1}^N S^{2n}_{iz},
\label{eq:H}
\end{equation}
where $n>0$ is an integer. The spin components satisfy 
\begin{equation}
\{S_{i\alpha},S_{j\gamma}\}=\delta_{ij}\epsilon_{\alpha \gamma
\delta}S_{j\delta},
\label{eq:spin-PB}
\end{equation}
where $\epsilon_{\alpha\gamma\delta}$ is the fully antisymmetric
Levi-Civita symbol. Here and in the following, we use Roman indices to
label the spins and Greek indices to denote the spin components. Noting that the canonical variables for the $i$-th spin are $\cos \theta_i$ and
$\phi_i$, the Poisson bracket $\{,\}$ appearing in
Eq.~(\ref{eq:spin-PB}) is defined for two functions $A,B$ of the spins by~\cite{Gupta:2011}
$\{A,B\}\equiv \sum_{i=1}^N (\partial A/\partial \phi_i\,\partial
B/\partial (\cos \theta_i)-\partial A/\partial (\cos \theta_i)\,\partial B/\partial \phi_i)$, which may be re-expressed
as 
\begin{equation}
\{A,B\}=\sum_{i=1}^N {\bf S}_i \cdot \frac{\partial A} {\partial
{\bf S}_i}\times \frac{\partial B}{\partial {\bf S}_i}.
\label{eq:poisson-bracket}
\end{equation}

We now explain the various terms appearing in Eq.~(\ref{eq:H}). Here, the first term with $J > 0$ on the right hand side models a ferromagnetic mean-field
interaction between the spins. On the other hand, the second term with
$D>0$ on the right hand side accounts
for local anisotropy: for example, $n=1$ (respectively, $n=2$) models quadratic
(respectively, quartic) anisotropy, and will be referred to below as the
quadratic (respectively, the quartic) model. Single-spin Hamiltonian of 
Heisenberg spins and involving quadratic and quartic terms has been considered previously in the
literature, see, e.g., Ref.~\cite{Kventsel:1984}. The anisotropy term in
Eq.~(\ref{eq:H}) lowers energy by having the
magnetization vector 
\begin{equation}
{\bf m}\equiv\frac{\sum_{i=1}^N{\bf S}_i}{N}=(m_x,m_y,m_z)
\end{equation}
pointing in the $xy$
plane. The length of the magnetization vector is given by
$m \equiv \sqrt{m_x^2+m_y^2+m_z^2}$.
The coupling constant $J$ in Eq.~(\ref{eq:H}) has been scaled down by
the system size $N$ to order to make the
energy extensive, in accordance with the Kac
prescription \cite{Kac:1963}. The system~(\ref{eq:H}) is however
intrinsically non-additive, since extensivity does not guarantee
additivity, although the converse is true. In the following, we set $J$ to unity
without loss of generality, and also take unity for the Boltzmann constant. 

In dimensionless time, the dynamics of the system (\ref{eq:H}) is governed by the set of
coupled first-order differential equations~\cite{Gupta:2011}
\begin{equation}
\dot{{\bf S}}_i=\{{\bf S}_i,H\}; ~~~~i=1,2,\ldots,N,
\label{eq:eom}
\end{equation}
where the dot denotes derivative with respect to time. Using Eq.~(\ref{eq:eom}), we obtain the dynamical evolution of the spin
components as
\begin{eqnarray}
\label{eq:eom-Six}
\dot{S}_{ix} &=& S_{iy}m_z-S_{iz}m_y - 2nDS_{iy}S^{2n-1}_{iz}, \\
\label{eq:eom-Siy}
\dot{S}_{iy} &=& S_{iz}m_x-S_{ix}m_z + 2nDS_{ix}S^{2n-1}_{iz}, \\
\label{eq:eom-Siz}
\dot{S}_{iz} &=& S_{ix}m_y-S_{iy}m_x.
\end{eqnarray}
Taking the vector dot product of both sides of Eq.~(\ref{eq:eom}) with ${\bf
S}_i$, it is easily seen that the dynamics conserves the length of each
spin. Summing Eq.~(\ref{eq:eom-Siz}) over $i$, we find that
$m_z$ is a constant of motion. Note that for the special case of no
anisotropy ($D=0$), the total magnetization $m$ is a constant of
motion. The total energy of the system is a constant of motion
under the dynamical evolution~(\ref{eq:eom}), and as such, the latter models microcanonical dynamics of the
system~(\ref{eq:H}). We remark that the dynamical setting of
Eq.~(\ref{eq:eom}) is very different from that involving particles
characterized by generalized coordinates and momenta and time evolution governed by a Hamiltonian given by a sum of a kinetic and a potential
energy contribution, e.g., that of the celebrated Hamiltonian
mean-field~(HMF)
model~\cite{Campa:2014}. As a result, none of the results, static or
dynamic, derived for the latter may be a priori expected to apply to the
model~(\ref{eq:H}). 
From Eqs.~(\ref{eq:eom-Six})-(\ref{eq:eom-Siz}), we obtain the time evolution of the variables $\theta_i$ and $\phi_i$ as
\begin{eqnarray}
\dot{\theta}_i &=& m_x\sin\phi_i - m_y \cos\phi_i, \label{eq:eom-theta}
\\
\dot{\phi}_i &=& m_x \cot\theta_i \cos\phi_i + m_y \cot\theta_i
        \sin\phi_i - m_z+ 2nD\cos^{2n-1}\theta_i. 
\label{eq:eom-phi}
\end{eqnarray}

Equations~(\ref{eq:eom-Six})-(\ref{eq:eom-Siz}) may be interpreted as
the precessional dynamics of the spins in an effective magnetic field ${\bf
h}^{\rm eff}_i\equiv {\bf h}^{\rm eff}_i(\{{\bf S}_i\})$:
\begin{equation}
\dot{\bf S}_i={\bf S}_i \times {\bf h}^{\rm eff}_i,
\label{eq:eom-precessional}
\end{equation}
where ${\bf h}^{\rm eff}_i$, the effective field for the $i$-th spin, is obtained from the Hamiltonian (\ref{eq:H})
as
\begin{equation}
{\bf h}^{\rm eff}_i = -\frac{\partial H}{\partial {\bf S}_i}={\bf m}+{\bf
h}^{\rm aniso}_i;~~{\bf h}^{\rm aniso}_i \equiv (0,0,-2nDS_{iz}^{2n-1}).
\end{equation}
Thus, the effective magnetic field has a global and a local
contribution, with the former being due to the magnetization set up in
the system by the effect of all the spins, and the latter being due to the field ${\bf
h}^{\rm aniso}_i$ set up for individual spins by the anisotropy
term in the Hamiltonian (\ref{eq:H}).

The paper is organized as follows. In Section~\ref{sec:previous}, we
summarize previous studies of model~(\ref{eq:H}) for $n=1$, which is
followed in Section~\ref{sec:queries} by listing of our queries in this
work, namely, the relaxation properties of the deterministic
dynamics~(\ref{eq:eom}) and the corresponding stochastic dynamics constructed
in the spirit
of the stochastic Landau-Lifshitz-Gilbert (LLG) equation~\cite{Lakshmanan:2011}. Here, we
also give a summary of results obtained in this work. The following sections are then
devoted to a derivation of these results. We start with a derivation of
the equilibrium
properties of the model~(\ref{eq:H}) in 
Section~\ref{sec:equilibrium-properties}. This is followed in Section~
\ref{sec:noiseless} by a study of the deterministic
dynamics~(\ref{eq:eom}) in the limit $N \to \infty$ in terms of the
so-called Vlasov equation in Subsection~\ref{sec:noiseless-infiniteN}. Here we also study linear stability of a
representative stationary state of the Vlasov equation, and demarcate for two
representative values of $n$ (namely, $n=1,2$) regions
in the parameter space where the state is stable under the Vlasov
evolution. Subsection~\ref{sec:noiseless-finiteN} is devoted to a
discussion on the behavior of the dynamics when $N$ is large but finite. The behavior
 of the stochastic dynamics in the limit $N \to \infty$ and in the case when
 $N$ is large but finite are taken up in Section~\ref{sec:noisy}. In
 this section, we also discuss a Monte Carlo scheme that serves as an
 alternative to the stochastic LLG scheme to study effects of noise on
 the deterministic dynamics~(\ref{eq:eom}). 
All throughout, we provide numerical checks of our
theoretical predictions, considering 
$n=1,2$ in the Hamiltonian~(\ref{eq:H}). We draw our conclusions in
Section~\ref{sec:conclusions}. Some of the technical details of our
analytic and numerical analysis are
collected in the three appendices.

\section{Previous studies}  
\label{sec:previous}
The quadratic model was first considered in Ref.~\cite{Gupta:2011} that
addressed the equilibrium and relaxational properties of the model. The
system was shown to exhibit in Boltzmann-Gibbs microcanonical equilibrium a
magnetized (equilibrium magnetization $m_{\rm eq} \ne 0$) phase at low values of the energy $\epsilon$ per
spin and a nonmagnetized ($m_{\rm eq}=0$) phase at high values, with a continuous
transition between the two occurring at a critical value $\epsilon_c$. It
was established that within microcanonical dynamics
and for a
class of nonmagnetized initial states, there exists a threshold
energy $\epsilon^\star < \epsilon_c$, such that in the energy range
$\epsilon^\star < \epsilon < \epsilon_c$, relaxation to equilibrium
magnetized state occurs over a time that scales
superlinearly with $N$~\cite{Gupta:2011,Lourenco:2015}. On the other hand, for energies $\epsilon <
\epsilon^\star$, the dynamics shows a fast relaxation out of the initial
nonmagnetized state over a time that scales as logarithm of $N$. The
particular initial state that was considered was the so-called
waterbag (WB) state, in which the spins have $\phi$'s chosen
independently and uniformly over the interval $[0,2\pi)$ and the
$\theta$'s chosen independently and uniformly over an interval symmetric about
$\pi/2$, that is, over the interval $[\pi/2-a,\pi/2+a]$, with $a$
being a real positive quantity. These results, obtained on the basis of
numerical integration of the equations of motion, were complimented by
an analytical study in the limit $N \to \infty$ of the time evolution,
\`{a}
la a Vlasov-type equation, of the single-spin phase space distribution.
The distribution counts the fraction of the total number of spins that
have given $\theta$ and $\phi$ values. It was demonstrated that the
distribution associated to the WB state is stationary under the
Vlasov evolution, but is unstable for energies below $\epsilon^\star$ and stable for
energies above. For finite $N$, the eventual relaxation to equilibrium
observed for energies $\epsilon^\star < \epsilon < \epsilon_c$ was
accounted as due to statistical fluctuations adding non-zero
finite-$N$ corrections to the Vlasov equation that are at least of order
greater than $1/N$~\cite{Barre:2014}. The WB state that for energies
$\epsilon^\star < \epsilon < \epsilon_c$ is stationary and stable in an
infinite system but which shows a slow evolution for finite $N$
exemplifies the
so-called quasistationary states (QSSs)~\cite{Campa:2014}.

\section{Our queries and summary of results obtained}
\label{sec:queries}
Starting from the premises discussed in the previous section, we pursue in this work a 
detailed characterization of the relaxational dynamics and its ubiquity
in the context of long-range spin models, by considering the
model~(\ref{eq:H}) for general values of $n$. We study for general $n$
the Boltzmann-Gibbs microcanonical equilibrium properties of the model, deriving in particular an
expression for the continuous phase transition point $\epsilon_c(n)$, such that the system is in a magnetized phase for lower energies and in
a nonmagnetized phase for higher energies. Though not guaranteed for LRI
systems~\cite{Campa:2014}, by virtue of the model~(\ref{eq:H}) exhibiting a continuous
transition in equilibrium, we conclude by invoking established
results~\cite{Bouchet:2005} that microcanonical and canonical ensembles are equivalent in
equilibrium. Consequently, one may associate to every value of the
conserved microcanonical energy density $\epsilon$ a temperature $T$ of the system in canonical equilibrium that guarantees
that the average energy in canonical equilibrium equals
$\epsilon$ in the limit $N \to \infty$. This allows to also derive the
phase diagram of the model~(\ref{eq:H}) in canonical equilibrium. 

The WB single-spin
distribution is non-analytic at $\theta=\pi/2 \pm a$, and one may wonder
as to whether such a peculiar feature led to quasitationarity in the
$n=1$ model reported in Refs.~\cite{Gupta:2011,Barre:2014} and
summarized in the preceding section. As a
counterpoint, and to
demonstrate that quasistationarity is rather generic to the
model~(\ref{eq:H}), we consider as
initial states suitably smoothened versions of the WB state, the
so-called Fermi-Dirac (FD) state, for which the single-spin distribution
is a perfectly analytic function, and
study its evolution in time. A linear stability
analysis of the FD state under the infinite-$N$ Vlasov dynamics establishes the existence of a threshold
energy value $\epsilon^\star(n)$, such that the state is stationary but linearly unstable under the dynamics for energies $\epsilon <
\epsilon^\star(n)$. For finite $N$, we establish that the relaxation to
equilibrium occurs as a two-step process: an initial relaxation from the
FD state to a magnetized QSS, followed by a relaxation of the latter
over a timescale $\sim N$ to Boltzmann-Gibbs microcanonical equilibrium. The
magnetized QSS has thus a lifetime $\sim
N$. For energies
$\epsilon^\star(n) < \epsilon < \epsilon_c(n)$, however, the FD state is dynamically stable under the Vlasov evolution, exhibiting for finite
$N$ a relaxation towards equilibrium over a scale $\sim N^\alpha$,
where the exponent $\alpha$ has an essential dependence on $n$. In this
case, one concludes observing a nonmagnetized QSS with a lifetime $\sim
N^\alpha$. As for $\alpha$, while
one obtains for $n=1$ the value $\alpha=3/2$ (as opposed to the value
$\alpha=2$ for the WB state reported in Ref.~\cite{Lourenco:2015}), one observes a linear
dependence ($\alpha=1$) for the quartic model. Note
that the two-step relaxation process for energies $\epsilon <
\epsilon^\star(n)$ was not reported in previous studies of the model
(\ref{eq:H}), see Refs.~\cite{Gupta:2011,Barre:2014}, and is being
reported here for the first time.  While magnetization $m$
turns out to be a useful macroscopic observable to monitor in order to establish the aforementioned relaxation scenario, it does not serve
the purpose when considering energies $\epsilon >
\epsilon_c(n)$ where both the initial FD and
the final Boltzmann-Gibbs microcanonical equilibrium state are nonmagnetized. Here, by identifying a suitable
observable (e.g., $\sum_{i=1}^N \cos^4 \theta_i/N$ and $\sum_{i=1}^N
\cos^2 \theta_i/N$ for respectively the quadratic and the quartic
model), we show that the relaxation to equilibrium occurs over a timescale that has an $N$ dependence distinct from what
was observed for magnetization relaxation for energies
$\epsilon^\star(n) < \epsilon < \epsilon_c(n)$. Namely, the relaxation
time scales as $N^2$ for the quadratic model and as $N^{3/2}$ for the
quartic model. We may thus conclude for energies $\epsilon >
\epsilon_c(n)$ the existence of a nonmagnetized QSS with a lifetime that
diverges with the system size. We stress that the existence of QSSs with
lifetimes $\sim N^2$ was not discussed in previous studies~\cite{Gupta:2011,Barre:2014}, and it is here
that we report on such states for the first time.

Our next issue of investigation is the robustness of QSSs with respect to
fluctuations induced through contact with the external environment.
Modelling the environment as a heat bath, previous studies of Hamiltonian particle dynamics (e.g., that of the HMF
model) have invoked a scheme of coupling to the environment that allows
for energy exchange and consequent stochastic Langevin evolution of the system. These studies have suggested
a fast relaxation to equilibrium over a size-independent timescale
provided the noise is strong
enough~\cite{Baldovin:2006-1,Baldovin:2006-2,Baldovin:2009}. In the
context of the model~(\ref{eq:H}), in order to assess the effects of
noise induced by the external environment, we study a stochastic version of
the dynamics~(\ref{eq:eom-precessional}) that considers the effective field ${\bf
h}_i^{\rm eff}$, see Eq.~(\ref{eq:eom-precessional}), to have an additional stochastic component due to 
interaction with the environment. The resulting
dynamics, built in the spirit
of the stochastic Landau-Lifshitz-Gilbert equation well known in
studies of dynamical properties of magnetic systems (see
Ref.~\cite{Lakshmanan:2011} for a review), reads
\begin{equation}
\dot{\mathbf{S}}_i= \mathbf{S}_i \times
\left(\mathbf{h}^{\rm eff}_i+\bfeta_i(t)\right) - \gamma \mathbf{S}_i
\times \left(\mathbf{S}_i \times \left(\mathbf{h}^{\rm eff}_i+\bfeta_i(t)
\right)\right),
\label{eq:eom-noise}
\end{equation}
where the second term on the right represents dissipation with the real
parameter $\gamma>0$ being the dissipation constant, and $\bfeta_i(t)$ is a Gaussian white noise with independent
components that satisfy
\begin{eqnarray}
&&\langle \eta_{i\mu} (t) \rangle=0,~\langle \eta_{i\mu}(t)\eta_{j\nu}(t')
\rangle=2\delta_{ij}\delta_{\mu\nu}{\cal D}\delta(t-t').
\end{eqnarray}
Here, ${\cal{D}}>0$ is a real constant that characterizes the strength of the
noise.
Note that the stochastic dynamics~(\ref{eq:eom-noise}) conserves the
length of each spin, as does the deterministic
dynamics~(\ref{eq:eom-precessional}). The former models dynamics within
a canonical ensemble for which the energy is not conserved during the
dynamical evolution, while, as already mentioned earlier, the latter models energy-conserving
microcanonical dynamics.

The presence of noise in Eq.~(\ref{eq:eom-noise}) makes the state of the
system at a given time, characterized by the set of values $\{{\bf
S}_i(t)\}$, to vary from one realization of the dynamics to another,
even when starting every time from the same initial condition.
Although Eq.~(\ref{eq:eom-noise}) has the flavor of Langevin dynamics, it is
different in that the noise and dissipation terms are incorporated in a
way that it has the desirable feature of keeping the length of each spin
to be unity at all times during the
dynamical evolution. Since the noise terms in Eq.~(\ref{eq:eom-noise})
depend on the state of the system, itself stochastic in nature, the
noise is said to be multiplicative in common parlance. As we argue later
in the paper, requiring the
dynamics~(\ref{eq:eom-noise}) to relax at long times to Boltzmann-Gibbs canonical
equilibrium at a given temperature $T$ 
fixes the constant ${\cal D}$ to be related to $\gamma$ in the manner
\begin{equation}
{\cal D}=\gamma T/(1+\gamma^2),
\end{equation}
a choice we also consider in this work. 
Our numerical simulation of the
dynamics~(\ref{eq:eom-noise}) follows the scheme detailed in
Appendix~C. The results show that in presence of strong-enough
noise, the system shows a fast relaxation to Boltzmann-Gibbs equilibrium on a
size-independent time scale, with no existence of intermediate
quasistationary states. We also implement a Monte Carlo
scheme as an alternative to the dynamics~(\ref{eq:eom-noise}) to study the effects of environment-induced noise on the
dynamics~(\ref{eq:eom}). On implementing the scheme, we find similar to the study of the
dynamics~(\ref{eq:eom-noise}) a fast relaxation to
Boltzmann-Gibbs equilibrium on a size-independent timescale. Our studies thus serve
to reaffirm what has been observed earlier in particle dynamics of LRI
systems, namely, that quasistationarity, observed in conservative
dynamics, is completely washed away in presence of stochasticity in the
dynamics.

\section{Equilibrium properties}
\label{sec:equilibrium-properties}

In this section, we investigate the properties of the system~(\ref{eq:H}) in the
thermodynamic limit $N \to \infty$ and in
canonical equilibrium at temperature $T=1/\beta$. Note that model~(\ref{eq:H}) is a
mean-field system that describes the motion of a spin moving in a
self-consistent mean-field generated by its interaction with all the
spins, with the single-spin Hamiltonian given by
\begin{eqnarray}
&&\fl h(\theta,\phi,m_x,m_y,m_z) \equiv -m_x \sin \theta \cos \phi - m_y
\sin \theta \sin \phi - m_z \cos \theta +D \cos^{2n} \theta. 
\label{eq:single-spin-hamiltonian}
\end{eqnarray}
Consequently, it is rather straightforward to write down exact expressions
for the average magnetization and the
average energy in equilibrium and in the thermodynamic limit. For
example, the single-spin equilibrium distribution is given by
\begin{equation}
f_{\rm eq}(\theta,\phi)\propto \exp(-\beta h(\theta,\phi,m_x^{\rm
eq},m_y^{\rm eq},m_z^{\rm eq})),
\end{equation}
with the equilibrium magnetization components $(m_x^{\rm eq},m_y^{\rm
eq},m_z^{\rm eq})$ obeying the self-consistent equation
\begin{equation}
(m_x^{\rm eq},m_y^{\rm eq},m_z^{\rm eq})=\frac{\int \int \sin \theta{\rm
d}\theta {\rm d}\phi~
(m_x^{\rm eq},m_y^{\rm eq},m_z^{\rm eq})~f_{\rm eq}(\theta,\phi)}{\int \sin \theta {\rm d}\theta {\rm
d}\phi~ f_{\rm eq}(\theta,\phi)}.
\label{eq:self}
\end{equation}
With $D>0$, so that the system orders in
the $xy$-plane, we may choose the ordering direction to be along $x$ without loss
of generality, yielding $m_x^{\rm eq} \ne 0, m_y^{\rm eq}=m_z^{\rm
eq}=0$. From Eq.~(\ref{eq:self}), we thus obtain for $m_{\rm eq} \equiv m_x^{\rm eq}$ the equation~\cite{Gupta:2011}
\begin{equation}
m_{\rm eq} = \frac{\int {\rm d}\theta {\rm d}\phi \sin^2\theta
\cos\phi ~e^{\beta  m_{\rm eq} \sin\theta \cos\phi - \beta D
\cos^{2n}\theta}}{\int {\rm d}\theta {\rm d}\phi \sin\theta  ~e^{\beta 
m_{\rm eq} \sin\theta \cos\phi - \beta D \cos^{2n}\theta}}.
\label{eq:mx-eqlbm-average}
\end{equation}
The average energy per spin equals~\cite{Gupta:2011}
\begin{equation}
\epsilon= - \frac{1}{2}  m_{\rm eq}^2 +
D \langle \cos^{2n}
\theta \rangle_{\rm eq},
\label{eq:energy-per-spineqlbm-average}
\end{equation}
with
\begin{equation}
\langle \cos^{2n} \theta \rangle_{\rm eq} = \frac{\int {\rm d}\theta {\rm
d}\phi \sin\theta \cos^{2n}\theta~e^{\beta  m_{\rm eq} \sin\theta \cos\phi - \beta D
\cos^{2n}\theta}}{\int {\rm d}\theta {\rm d}\phi \sin\theta  ~e^{\beta m_{\rm eq} \sin\theta \cos\phi - \beta D \cos^{2n}\theta}}.
\end{equation}

From the fact that the model~(\ref{eq:H}) with $n=1$ shows a continuous phase
transition in magnetization across critical inverse temperature
$\beta_c=1/T_c$~\cite{Gupta:2011}, we may anticipate that so is the case for general $n$. Consequently, we may consider Eq.~(\ref{eq:mx-eqlbm-average}) close
to the critical point, i.e., for $\beta \gae \beta_c$, when $m_{\rm eq}$ is small so that the equation
may be expanded to leading order in $m_{\rm eq}$, as
\begin{eqnarray}
&&\fl m_{\rm eq} \Big(\int {\rm d}\theta {\rm d}\phi
        \sin\theta ~e^{-\beta D \cos^{2n}\theta}-~\beta  \int {\rm d}\theta {\rm d}\phi \sin^3\theta \cos^2\phi
        ~e^{-\beta D \cos^{2n}\theta} \Big) = 0.
\end{eqnarray}
With $m_{\rm eq}\ne 0$, one obtains $\beta_c$ as the value of
$\beta$ that sets the bracketed quantity to zero; on performing the
integrals, one obtains $\beta_c$
to be satisfying
\begin{equation}
1 - \frac{2}{\beta_c} = \frac{  \Gamma(3/(2n)) -\Gamma(3/(2n), \beta_c
D)  }{ (\beta_c D)^{2/(2n)} \Big[ \Gamma(1/(2n)) -\Gamma(1/(2n), \beta_c D)  \Big]}.
\label{eq:beta_c}
\end{equation}
Here, $\Gamma(s)$ is the Gamma function, while $\Gamma(s,x)$ is the upper incomplete Gamma function. At the critical
point, when we have $m_{\rm eq}=0$, one obtains the critical energy density as $\epsilon_c
= D \langle \cos^{2n}\theta \rangle_{\rm eq} $, that is,
\begin{equation}
\epsilon_c = D\frac{\int {\rm d}\theta {\rm d}\phi
\sin\theta \cos^{2n} \theta~e^{ - \beta D
\cos^{2n}\theta}}{\int {\rm d}\theta {\rm d}\phi \sin\theta  ~e^{ - \beta D
        \cos^{2n}\theta}};
\end{equation}
on performing the integrals, we get
\begin{equation}
\epsilon_c=  \frac{ \Gamma(1/(2n)) - 2n\Gamma(1+1/(2n),\beta_c D) }{2n\beta_c
\Big[  \Gamma(1/(2n)) - \Gamma(1/(2n),\beta_c D)\Big]}.
\label{eq:ec}
\end{equation}
Note that for $n=1$, one may check using the above expressions that
$1-2/\beta_c=1/(2\beta_cD)-e^{-\beta_cD}/(\sqrt{\pi\beta_cD}
\mathrm{Erf}[\sqrt{\beta_cD}])$ and that
$\epsilon_c=D\Big(1-2/\beta_c\Big)$, where $\mathrm{Erf}[x] \equiv (2/\sqrt{\pi})\int_0^x {\rm d}t~ e^{-t^2}$ is the
error function, as was reported in Ref.~\cite{Gupta:2011}.

\begin{figure}[!ht]
\centering
\includegraphics[width=7cm]{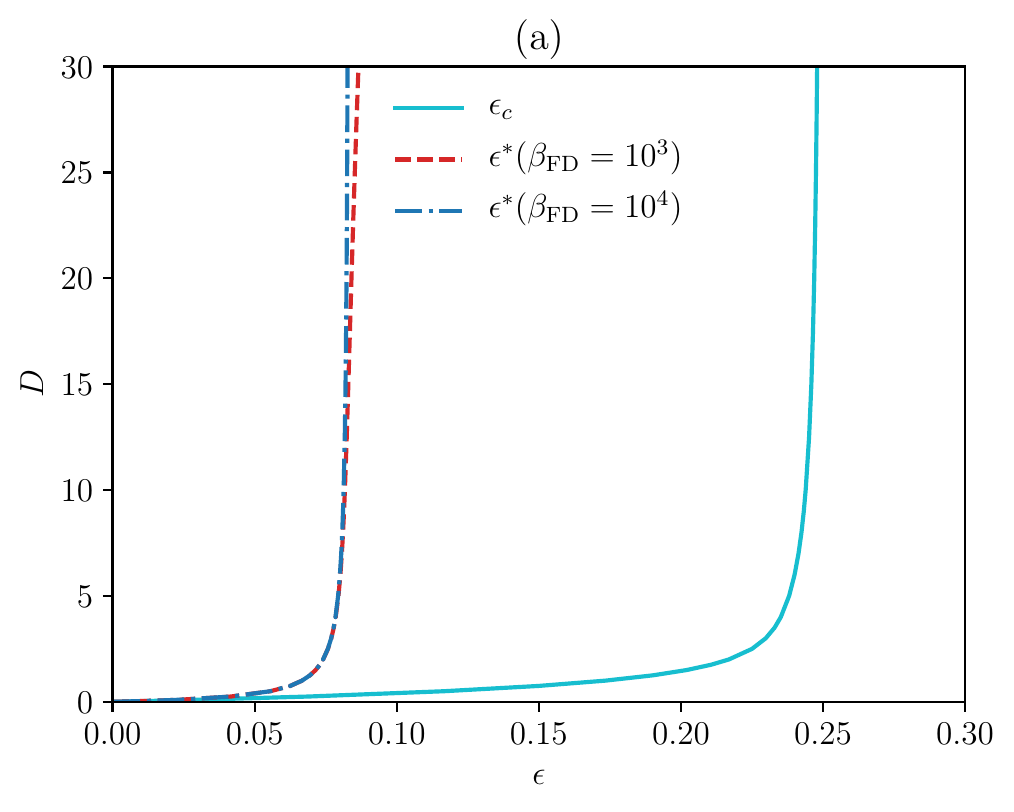}
\includegraphics[width=7cm]{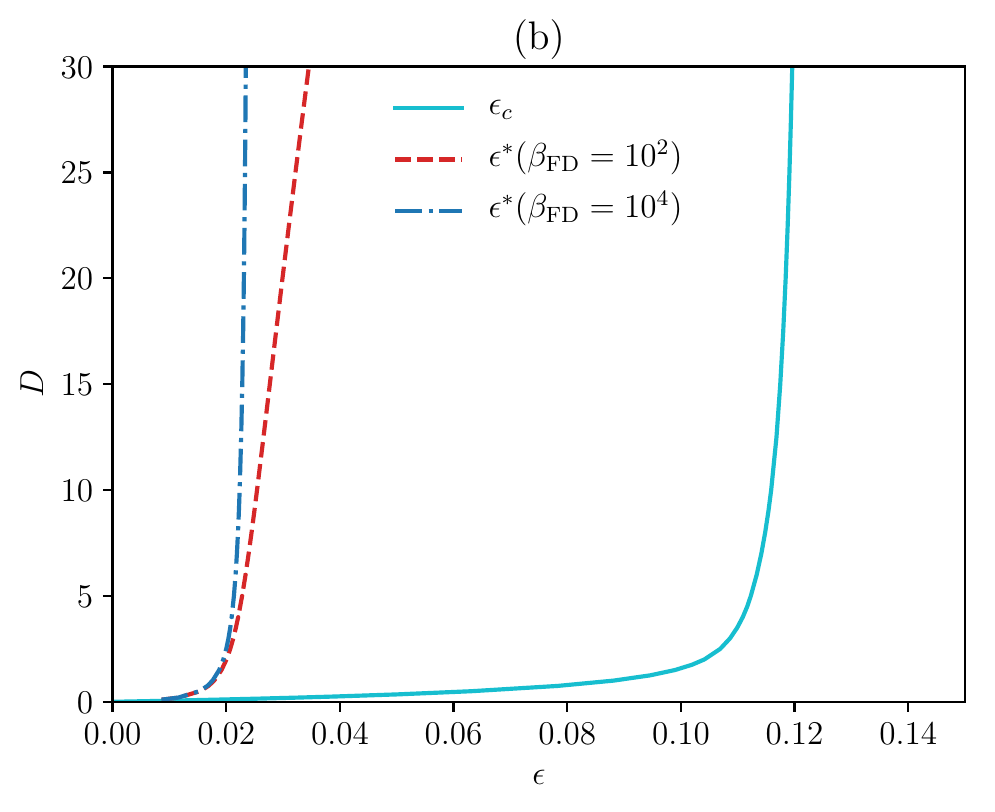}
\caption{Phase diagram of the model (\ref{eq:H}) for $n=1$ (left panel)
and $n=2$ (right panel), showing both the
equilibrium phase boundary $\epsilon_c$ and the Vlasov stability
boundary $\epsilon^\star$ corresponding to the FD state~(\ref{eq:FD})
for two values of $\beta_{\rm FD}$, for large $\beta_{\rm FD}$. Both for $n=1$ and $n=2$, the result for
the larger $\beta_{\rm FD}$ value coincides with that obtained for the WB state~(\ref{eq:WB-state}).}        
\label{fig:phase-diagram}
\end{figure}

\begin{figure}[!ht]
\centering
\includegraphics[width=7cm]{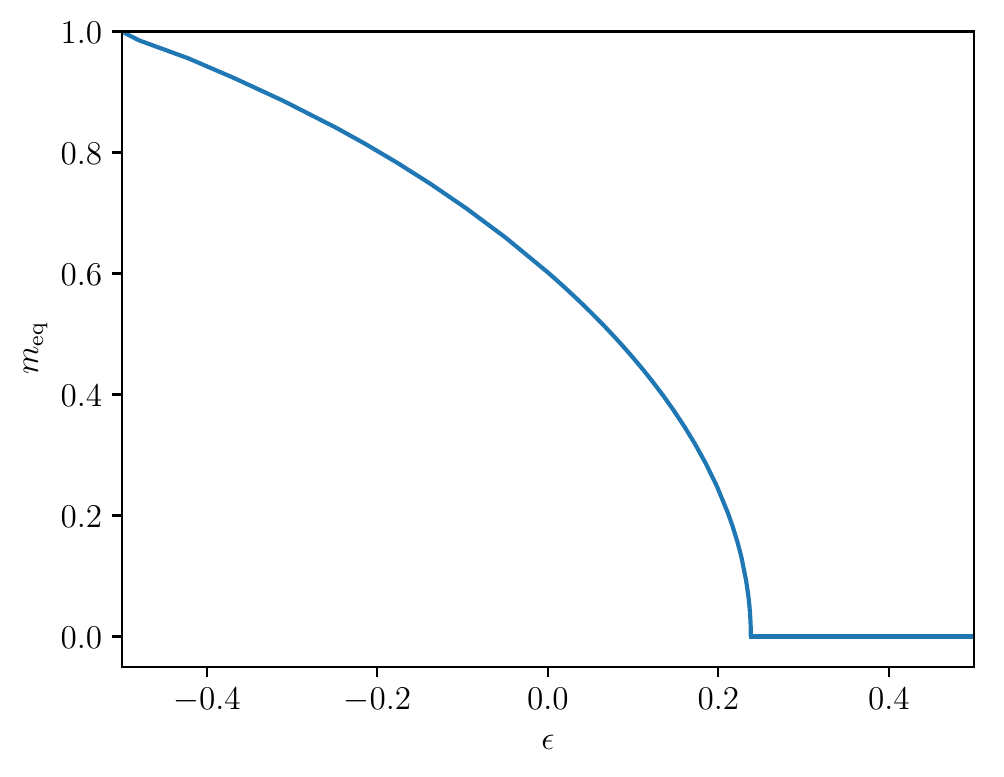}
\includegraphics[width=7cm]{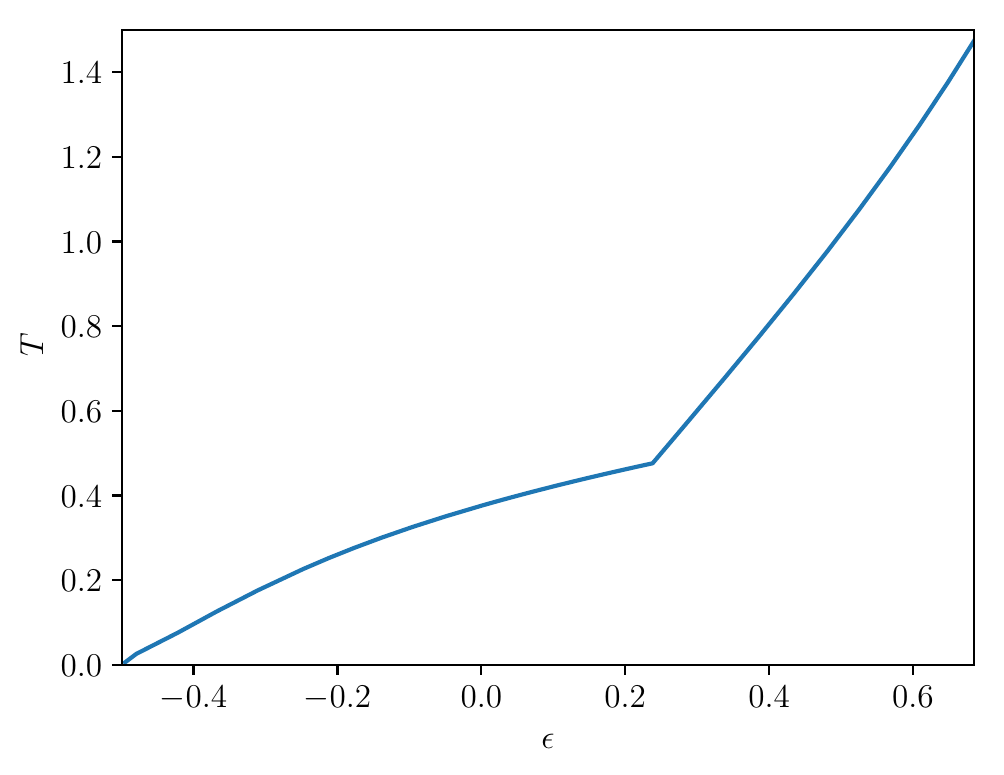}
\caption{Magnetization $m_{\rm eq}$ and temperature $T$ as a function of
energy density $\epsilon$ in microcanonical equilibrium for the
model~(\ref{eq:H}) with $n=1$ and $D=5.0$. The
magnetization decreases continuously from unity to zero at the critical
energy density $\epsilon_c$, obtained from Eq.~(\ref{eq:ec}) as
$\epsilon_c \approx 0.2381$, and remains zero
at higher energies. Correspondingly, the $T$ versus $\epsilon$ curve (the
so-called caloric curve) shows a discontinuity at the critical energy
$\epsilon_c$, namely, ${\rm d}\epsilon/{\rm d}T|_{\epsilon_c-} \ne{\rm
d}\epsilon/{\rm d}T|_{\epsilon_c+}$}.
\label{fig:eTm}
\end{figure}

Since the phase transition exhibited by the model~(\ref{eq:H}) is a
continuous one, the canonical and microcanonical ensemble properties in
equilibrium would be
equivalent~\cite{Bouchet:2005}, and hence, Eq.~(\ref{eq:ec}) also gives
the conserved microcanonical energy density at the transition point.
Figure~\ref{fig:phase-diagram} shows for $n=1,2$ the energy density $\epsilon_c$ as
a function of $D$, obtained by first solving numerically for a given $D$ the
transcendental equation~(\ref{eq:beta_c}) for $\beta_c$ and then using
the obtained value of $\beta_c$ in Eq.~(\ref{eq:ec}). Moreover, one may
construct a one-to-one mapping between a value of microcanonical
equilibrium energy density $\epsilon$ and canonical equilibrium temperature $T$ by first solving
Eq.~(\ref{eq:mx-eqlbm-average}) at a given $T$ to obtain the equilibrium
magnetization $m_{\rm eq}$, then substituting in
Eq.~(\ref{eq:energy-per-spineqlbm-average}) to obtain the corresponding
average energy in canonical equilibrium, and finally demanding that the
latter is the conserved energy density in microcanonical equilibrium.
On carrying out this program for $n=1$ and $D=5.0$, one obtains the results shown in
Fig.~\ref{fig:eTm}, where we also show $m_{\rm eq}$ as a function of
microcanonical energy density $\epsilon$.

\section{Analysis of the deterministic
dynamics~(\ref{eq:eom-precessional})}
\label{sec:noiseless}

\subsection{Behavior in the limit $N \to \infty$}
\label{sec:noiseless-infiniteN}

We now discuss how the model~(\ref{eq:H}) in the limit $N \to \infty$
and under the dynamical
evolution~(\ref{eq:eom-precessional}) relaxes to equilibrium while
starting far from it, by invoking the corresponding Vlasov equation. The relaxation may be
characterized by monitoring the time evolution of the single-spin
distribution function $P_0({\bf S},t)$ normalized as $\int {\rm d}{\bf
S}~P_0({\bf S},t)=1~\forall~t$. As detailed in Appendix~B,
the time evolution of $P_0({\bf S},t)$ follows the Vlasov equation
\begin{eqnarray}
&& 
\frac{\partial P_0({\bf S},t)}{\partial
t} +\frac{\partial}{\partial{\bf
 S}}\cdot({\bf S}\times{\bf h}^{{\rm
 eff},0})P_0=0,
\label{eq:Vlasov-1}
\end{eqnarray}
where we have
\begin{equation}
\fl {\bf h}^{{\rm eff},0}\equiv {\bf h}^{{\rm eff},0}[P_0]={\bf
m}[P_0]+(0,0,-2nDS_z^{2n-1});~~{\bf m}[P_0]\equiv\int {\rm d}{\bf
S}~{\bf S}P_0({\bf
S},t).
\end{equation}

For later purpose, it is convenient to consider the single-spin
distribution $f(\theta,\phi,t)$, defined such that $f(\theta,\phi,t)\sin \theta {\rm d}\theta {\rm d}\phi$ is the
probability to have a spin at time $t$ with its angles between $\theta$ and
$\theta+{\rm d}\theta$ and between $\phi$ and $\phi+{\rm d}\phi$, and
which is related to $P_0({\bf S},t)$ as $f(\theta,\phi,t)=P_0({\bf S},t)$,
with the normalization $\int_0^{2\pi} {\rm d}\phi \int_0^\pi {\rm
d}\theta~\sin \theta~f(\theta,\phi,t)=1~\forall~t$. To obtain the time evolution of
$f$, let us express the second term on the left hand side of
Eq.~(\ref{eq:Vlasov-1}), which equals $({\bf S}\times{\bf h}^{{\rm
 eff},0})\cdot \partial P_0/\partial {\bf S}$, in spherical polar coordinates with unit vectors
$(\hat{r},\hat{\theta},\hat{\Phi})$ and 
$r=\sqrt{S^2_x+S^2_y+S^2_z}=1$. We get
\begin{eqnarray}
&&\fl \Big[(S_yh^{\rm eff}_z -S_zh^{\rm eff}_y) ( \cos\theta \cos\phi
~\hat{\theta} -\sin\phi ~ \hat{\phi}  )+ (S_zh^{\rm eff}_x -S_xh^{\rm
eff}_z) ( \cos\theta \sin\phi ~\hat{\theta} +\cos\phi ~\hat{\phi}  )  \nonumber \\
&&\fl - (S_xh^{\rm eff}_y -S_yh^{\rm eff}_x)  \sin\theta ~\hat{\theta}
\Big] \cdot \Big[   \hat{\theta}  \frac{\partial}{\partial \theta}    +
\frac{1}{\sin\theta} \hat{\phi} \frac{\partial}{\partial \phi}   \Big] f(\theta,\phi,t) \nonumber \\
&&\fl=  \Big[\cos\theta\cos\phi (S_yh^{\rm eff}_z -S_zh^{\rm eff}_y) + \cos \theta\sin\phi (S_zh^{\rm eff}_x -S_xh^{\rm eff}_z)-\sin\theta (S_xh^{\rm eff}_y -S_yh^{\rm eff}_x) \Big] \frac{\partial f}{\partial \theta} \nonumber \\
 &&\fl + \Big[ -  \sin\phi (S_yh^{\rm eff}_z -S_zh^{\rm eff}_y) + \cos\phi (S_zh^{\rm eff}_x -S_xh^{\rm eff}_z) \Big] \frac{1}{\sin\theta}\frac{\partial f}{\partial \phi} \nonumber \\
&&\fl= -\Big[m_y \cos\phi - m_x \sin\phi \Big] \frac{\partial
f}{\partial \theta}  + \Big[m_x\cot\theta \cos \phi + m_y\cot\theta \sin
\phi - m_z + (2n) D \cos^{2n-1} \theta \Big]  \frac{\partial f}{\partial
\phi}. \nonumber \\ 
\end{eqnarray}
Substituting in Eq.~(\ref{eq:Vlasov-1}), we get for the time evolution of $f(\theta,\phi,t)$ the equation
\begin{eqnarray}
&& \fl \frac{\partial f}{\partial t} =\Big(m_y \cos \phi - m_x \sin\phi
\Big) \frac{\partial f}{\partial \theta}-\Big( m_x \cot\theta \cos\phi+ m_y\cot\theta \sin\phi - m_z+(2n)D\cos^{2n-1} \theta \Big)\frac{\partial
f}{\partial \phi}, \nonumber \\ 
\label{eq:Vlasov-equation}
\end{eqnarray}
with 
\begin{eqnarray}
&&\fl(m_x,m_y,m_z)\equiv (m_x,m_y,m_z)[f]=\int \sin \theta'{\rm d}\theta' {\rm d}\phi'
(\sin \theta'\cos \phi',\sin \theta' \sin \phi',\cos
\theta')f(\theta',\phi',t). \nonumber \\ 
\end{eqnarray}

Let us consider as a far-from-equilibrium initial condition a state
$f_0(\theta,\phi)$ that is uniform in $\phi$ over $[0,2\pi)$ and uniform in $\theta$
over a symmetric interval about $\theta=\pi/2$:
\begin{eqnarray}
f_0(\theta, \phi) &=&
\frac{A}{2\pi}p(\theta),
\label{eq:f0}
\end{eqnarray}
with $p(\pi/2-\theta)=p(\pi/2+\theta)$.
Such a state is
evidently nonmagnetized, i.e., with $m_x=m_y=m_z=0$. It
then follows that such a state is a stationary solution of the Vlasov
equation~(\ref{eq:Vlasov-equation}). We now study dynamical stability of
this stationary state with respect of
fluctuations. 
The method of
analysis follows closely the one pursued in Ref.~\cite{Gupta:2011}. To this end, we linearize the Vlasov
equation~(\ref{eq:Vlasov-equation}) with respect to small fluctuations
$f_1(\theta,\phi,t)$ by expanding $f(\theta,\phi,t)$ as
\begin{equation}
f(\theta,\phi,t)=f_0+\lambda f_1(\theta,\phi,t),
\label{eq:f-expansion}
\end{equation} 
with $|\lambda| \ll 1$. 
The linearized Vlasov equation reads 
\begin{equation}
\frac{\partial f_1}{\partial t}=\Big(\widetilde{m}_y \cos \phi -
\widetilde{m}_x \sin\phi \Big) \frac{\partial f_0}{\partial \theta} -
(2n)D\cos^{2n-1}\theta \frac{\partial f_1}{\partial \phi},
\label{eq:Vlasov-linearized}
\end{equation}
where $\widetilde{m}_x$ and $\widetilde{m}_y$ are linear in $f_1$:
$(\widetilde{m}_x,\widetilde{m}_y)[f_1]\equiv\int {\rm d}\theta {\rm d}\phi
\sin\theta~(\sin \theta\cos\phi,\sin \theta\sin \phi)
~f_1(\theta,\phi,t)$.

Now, since $f_1(\theta,\phi,t)$ is $2\pi$-periodic in $\phi$, we may
implement the following Fourier expansion:
\begin{equation}
f_1(\theta,\phi,t) = \sum_{k} \int {\rm d}\omega~g_k(\theta,\omega)
e^{i(k\phi + \omega t)}.
\end{equation}
In the long-time limit, we may expect the linearized Vlasov dynamics to be dominated by
the Fourier mode of frequency $\omega$ with the smallest imaginary part,
so that one effectively has $f_1(\theta,\phi,t) = \sum_{k}
~g_k(\theta,\omega) e^{i(k\phi + \omega t)}$, yielding
\begin{equation}
\widetilde{m}_x=\pi e^{i\omega t} \big(I_{+} +
I_{-}\big),~\widetilde{m}_y=i\pi e^{i\omega t} \big(I_{+} - I_{-}\big),
\label{eq:m-Fourier-expansion}
\end{equation}
with $I_{\pm}=\int {\rm d}\theta \sin^2\theta ~ g_{\pm 1} (\theta,
\omega)$. It then follows that the relevant
eigenmodes of Eq.~(\ref{eq:Vlasov-linearized}) are those with $k= \pm 1$. Indeed, as follows from Eq.~(\ref{eq:Vlasov-linearized}), modes $k \ne
\pm 1$ only oscillate in time. This fact that only the long-wavelength
(i.e., small-$k$) mode perturbations are the ones that determine the
stability of stationary states holds in general for long-range systems.
In this regard, the reader may refer to the phenomenon of Jeans
instability in a prototypical long-range system, the gravitational systems~\cite{book-Jeans}.

Using
Eq.~(\ref{eq:m-Fourier-expansion}) and the aforementioned expansion of
$f_1$ in Eq.~(\ref{eq:Vlasov-linearized}), we find that the coefficients
$g_{\pm 1}(\theta,\omega)$ satisfy
\begin{equation}
g_{\pm 1}(\theta, \omega)=\pi\frac{\partial f_0}{\partial \theta}
~\frac{I_{\pm}}{(2n)D\cos^{2n-1}\theta \pm \omega}.
\end{equation}
Multiplying both sides by $\sin^2 \theta$ and then integrating over
$\theta$, we find, by using the definition of the quantity $I_\pm$ and
the fact that $I_\pm \ne 0$, that 
\begin{equation}
I \equiv \pi  \int {\rm d}\theta ~\frac{\partial f_0}{\partial \theta}
~\frac{\sin^2 \theta}{(2n)D\cos^{2n-1}\theta \pm \omega} = 1.
\label{eq:stability-condition}
\end{equation}

\begin{figure}[!ht]
\centering
\includegraphics[width=7cm]{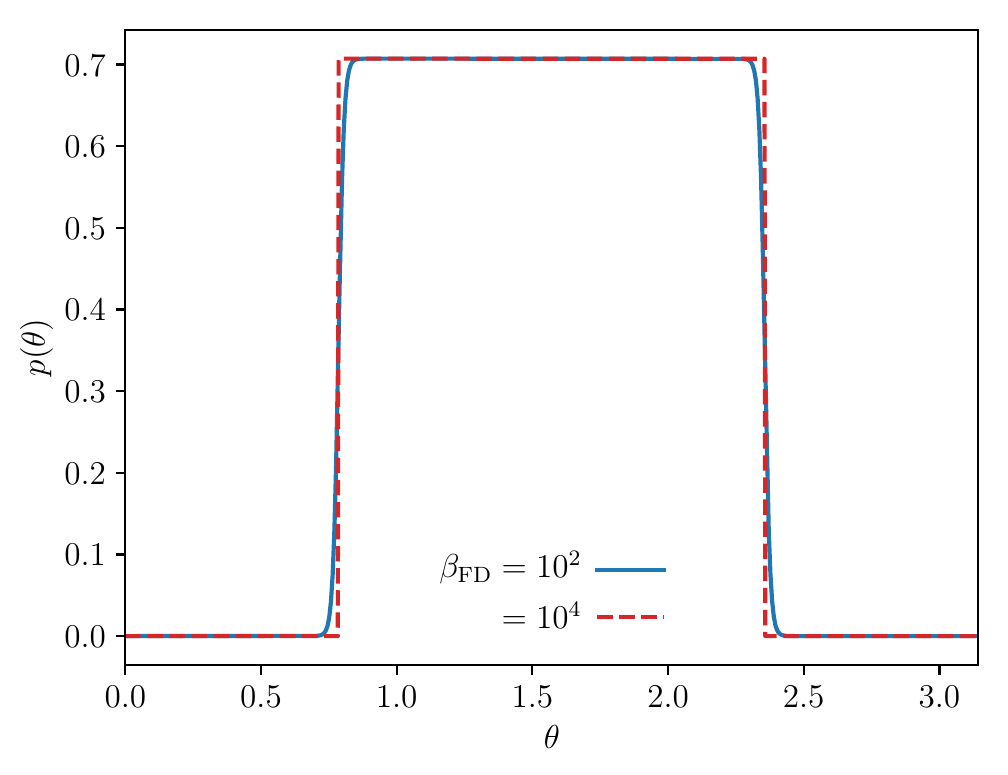}
\caption{The $\theta$-distribution $p(\theta)$, corresponding to the Fermi-Dirac (FD)
distribution~(\ref{eq:FD}), for two large values of $\beta_{\rm FD}$ and with $\mu=0.5$.}
\label{fig:ptheta}
\end{figure}

Let us consider as a representative example for $f_0(\theta,\phi)$ the
form
\begin{eqnarray}
f_0(\theta, \phi) &=&
\frac{A}{2\pi}p(\theta);~p(\theta)=\frac{1}{1+e^{\beta_{\rm FD}(\cos^2\theta - \mu)}},
\label{eq:FD}
\end{eqnarray}
where $\mu \equiv \sin^2a$ with $a>0$ and $\beta_{\rm FD} >0$ being real
parameters, and $A$ is the normalization constant. In the
limit $\beta_{\rm FD} \to \infty$, it is easy to see that $p(\theta)$ is a
uniform distribution over the range $\theta \in [\pi/2-a,\pi/2+a]$;
correspondingly, the distribution~(\ref{eq:FD}) becomes
\begin{eqnarray}
&&f_0(\theta,\phi)\nonumber \\
&&=\left\{
\begin{array}{ll}
               \frac{1}{2\pi}\frac{1}{2\sin a} & \mbox{if $\theta \in
               \left[\frac{\pi}{2}-a,\frac{\pi}{2}+a\right]$}, \phi \in
               [0,2\pi)\nonumber \\
               & \\
               0 & \mbox{otherwise},
               \end{array}
        \right. \\ 
\label{eq:WB-state}        
\end{eqnarray}
and is thus identical to the WB state~\cite{Gupta:2011}. For
finite but large $\beta_{\rm FD}$, the distribution is smoothened around the boundaries at
$\theta = \pi/2 \pm a$. Figure~\ref{fig:ptheta} shows $p(\theta)$ for
different values of $\beta_{\rm FD}$ and for $\mu=0.5$, which makes it evident the similarity in
the form of $p(\theta)$ to the Fermi-Dirac (FD) distribution.
Henceforth, we will refer to the distribution~(\ref{eq:FD}) as the FD
state. While it is not possible to derive analytical results for
the FD distribution for general $\beta_{\rm FD}$, simplifications occur
for large $\beta_{\rm FD}$ when exact expressions may be derived, as detailed below.

In Appendix~A, we show that for large $\beta_{\rm FD}$, we have
\begin{equation}
A = \frac{1}{2\sqrt{\mu}} \Big[ 1+ \frac{\pi^2}{24\beta_{\rm FD}^2 \mu^2} \Big], 
\label{eq:FD-norma}
\end{equation}
 correct to order $1/\beta_{\rm FD}^2$, while to same order, the energy corresponding to the state~(\ref{eq:FD}) is given by 
\begin{equation}
\epsilon = \frac{D}{2n+1} \Big[\mu^{2n/2} +
\frac{(2n)^2\pi^2}{24\beta_{\rm FD}^2} \mu^{(2n-4)/2} \Big]. 
\label{energy}
\end{equation}
Next, using Eq.~(\ref{eq:FD}) in 
Eq.~(\ref{eq:stability-condition}), we get to order $1/\beta_{\rm FD}^2$ the equation
\begin{equation}
g(\mu)\mu^{-1/2}+  \frac{\pi^2}{24\beta_{\rm FD}^2} \Big[ g(\mu) \mu^{-5/2} + 4
g''(\mu) \mu^{-1/2} \Big] = \frac{1}{nD},
\label{eq:spin4-FD-stability-criterion}
\end{equation}
where we have 
\begin{equation}
g(x) \equiv \frac{x^{(2n-1)/2}-x^{(2n+1)/2}}{(2n)^2D^2x^{2n-1}-\omega^2},
\end{equation}
while $\mu$ is to be obtained by solving Eq.~(\ref{energy}). The latter
equation gives for $n=1$ two possible values of $\mu$ given by
\begin{eqnarray}
\mu = \Big[ \frac{3\epsilon}{D} - \frac{\pi^2 D}{18\beta_{\rm FD}^2 \epsilon}
\Big] \hspace{0.25cm} {\rm and} \hspace{0.25cm} \frac{\pi^2
D}{18\beta_{\rm FD}^2 \epsilon},
\label{eq:roots-s2}
\end{eqnarray}
and for $n=2$ a single value given by
\begin{eqnarray}
\mu = \sqrt{\frac{5\epsilon}{D} - \frac{2\pi^2}{3\beta_{\rm FD}^2}}.
\label{eq:roots-s4}
\end{eqnarray}
Using Eqs.~(\ref{eq:roots-s2}) and (\ref{eq:roots-s4}) in
Eq.~(\ref{eq:spin4-FD-stability-criterion}) and retaining terms up to
order $1/\beta_{\rm FD}^2$, we finally obtain a relation connecting
$\omega,D,\beta_{\rm FD}$ and $\epsilon$, which is correct to same order. For $n=1$, we obtain two equations:
\begin{eqnarray}
\fl && D\omega^4 - 3\omega^4 \epsilon +D^2 \omega^2 \epsilon - 6
D^2\omega \epsilon + \frac{\pi^2 D}{18\beta_{\rm FD}^2\epsilon} \Big(
48D^4\epsilon-8D^3\omega+6D^2\omega\epsilon-18D^2\omega^2\epsilon-D\omega^4
\Big) \nonumber \\
\fl &&= \frac{2\pi^2 D^3 \omega^4}{3\beta_{\rm FD}^2 \epsilon} - \omega^6,
\label{eq:n2-omega-eqn2}
\end{eqnarray}
and
\begin{eqnarray}
&&\fl D \big( \widetilde{\alpha} -\omega^2 \big) \Big[  (1-\widetilde{\gamma})
\big( \widetilde\alpha-\omega^2 \big) + \frac{\pi^2}{24\beta_{\rm FD}^2
\widetilde\gamma^2} \Big( 11\widetilde\alpha \widetilde\gamma -
7\widetilde\alpha -3\widetilde\gamma \omega^2 - \omega^2 \Big) \Big] \nonumber \\
&& \fl + \frac{\pi^2}{24\beta_{\rm FD}^2 \widetilde\gamma^2} \Big[
\widetilde\alpha \widetilde\gamma +3\widetilde\alpha -6 \widetilde\alpha
\widetilde\gamma
\omega^2 + 6\widetilde\alpha \omega^2 -3\widetilde\gamma \omega^4 - \omega^4
\Big] = \big( \widetilde\alpha -\omega^2 \big)^3 -
\frac{2\pi^2D^2}{\beta_{\rm FD}^2 \widetilde\gamma} \big(
\widetilde\alpha -\omega^2 \big)^2, \nonumber \\
\label{eq:n2-omega-eqn1}
\end{eqnarray}
where we have $\widetilde\gamma \equiv 3\epsilon/D$ and $\widetilde\alpha \equiv
4D^2\widetilde\gamma$.  For $n=2$, Eqs.~(\ref{eq:roots-s4})
and~(\ref{eq:spin4-FD-stability-criterion}) give to order $1/\beta_{\rm
FD}^2$ a single equation:
\begin{eqnarray}
&& \fl 2D (\alpha-\omega^2) \Big[ (\gamma-\gamma^2) (\alpha-\omega^2)
-\frac{\pi^2}{24\beta_{\rm FD}^2 \gamma} \Big( 55\alpha - 63\gamma \alpha +15\gamma \omega^2 -7\omega^2 \Big) \Big] \nonumber \\
&& \fl +\frac{D\pi^2}{3\beta_{\rm FD}^2 \gamma} \Big[ -12D^2 \gamma^4 \alpha+
60D^2 \gamma^3 \alpha -16\gamma \omega^2 \alpha +16 \omega^2 \alpha
-\frac{15\omega^4 \gamma}{4} + \frac{3\omega^4}{4} \Big] \nonumber \\
&&\fl = (\alpha-\omega^2)^2 \Big[ \alpha-\omega^2 - \frac{3\pi^2
\alpha}{\beta_{\rm FD}^2 \gamma^2}\Big], \label{eq:n4-omega-eqn}
\end{eqnarray}
where we have $\gamma^2\equiv5\epsilon/D$ and
$\alpha\equiv16D^2\gamma^3$. On physical grounds, we would want
Eqs.~(\ref{eq:n2-omega-eqn2}), (\ref{eq:n2-omega-eqn1}),
and~(\ref{eq:n4-omega-eqn}) to be valid for all $\omega$, including
$\omega=0$. Equation~(\ref{eq:n2-omega-eqn2}) however gives for
$\omega=0$ an inconsistent relation $8\pi^2D^5/(3\beta_{\rm FD}^2)=0$, and hence, may be discarded. 

In the limit $\beta_{\rm FD} \to \infty$, when the FD state~(\ref{eq:FD}) reduces to the WB state~(\ref{eq:WB-state}), Eqs.~(\ref{eq:n2-omega-eqn1}) and~(\ref{eq:n4-omega-eqn}) reduce respectively to
\begin{eqnarray}
\omega^2 = 12D\epsilon + 3\epsilon - D,
\label{eq:omega-n2-beta-infinite}
\end{eqnarray}
and
\begin{eqnarray}
 \omega^2 = 2\sqrt{5\epsilon D} \Big[40\epsilon -1+ \sqrt{\frac{5\epsilon}{D}}\Big].
 \label{eq:omega-n4-beta-infinite}
\end{eqnarray}
Setting $\omega=0$ in Eqs.~(\ref{eq:omega-n2-beta-infinite})
and~(\ref{eq:omega-n4-beta-infinite}) yields the value
$\epsilon^\star(D,\beta_{\rm FD}\to \infty)$ of the energy density such that these equations give only real roots for the frequency $\omega$ for $\epsilon > \epsilon^*$. We find that $\epsilon^\star$ satisfies the following equations:
\begin{eqnarray}
\epsilon^\star &=& \frac{D}{3+12D} \hspace{10mm} {\rm for~} n=1,
\label{eq:n2-WB-estar}
\end{eqnarray}
and
\begin{eqnarray}
40\epsilon^\star + \sqrt{\frac{5\epsilon^\star}{D}} = 1 \hspace{10mm}
{\rm for~ } n=2.
\label{eq:n4-WB-estar}
\end{eqnarray}
 It then follows that the WB state is linearly unstable under the Vlasov
dynamics for energy density $\epsilon$ smaller than
$\epsilon^\star$, and is linearly stable for $\epsilon > \epsilon^\star$. For $\epsilon < \epsilon^\star$, the
perturbation $f_1(\theta,\phi,t)$ grows exponentially in time. Here, on setting $\omega^2=-\Omega^2$ with
real $\Omega$, one gets 
\begin{equation}
f_1(\theta,\phi,t) \sim e^{\pm i\phi + \Omega t}.
\label{eq:m-e<estar-0}
\end{equation}

For finite but large $\beta_{\rm FD}$, when
Eqs.~(\ref{eq:n2-omega-eqn1}) and~(\ref{eq:n4-omega-eqn}) are valid, we
may expect on the basis of the above that there exists an energy
threshold $\epsilon^\star(D,\beta_{\rm FD}~{\rm large})$ such that the
FD state is linearly unstable and that the
scaling~(\ref{eq:m-e<estar-0}) holds for energy
$\epsilon<\epsilon^\star$, while the state is stable for energies $\epsilon>\epsilon^\star$. Such an $\epsilon^\star$ may be obtained by setting $\omega=0$ in Eqs.~(\ref{eq:n2-omega-eqn1}) and~(\ref{eq:n4-omega-eqn}); there are more than one value of
$\epsilon^{\star}$ for given $D$ and $\beta_{\rm FD}$ that one obtains
in doing so, and we take for the physically meaningful $\epsilon^\star$
only the value that reduces to Eqs.~(\ref{eq:n2-WB-estar})
and~(\ref{eq:n4-WB-estar}) as one takes the limit $\beta_{\rm FD} \to
\infty$. The result for $\epsilon^\star$ as a function of $D$ is shown
in Fig.~\ref{fig:phase-diagram} for $n=1,2$.

\subsection{Behavior for finite $N$}
\label{sec:noiseless-finiteN}

Equation~(\ref{eq:Vlasov-1}) describes the time evolution in an
infinite system, and here we ask: what happens when the system size $N$
is large but finite? Such a situation arises while studying the
dynamics~(\ref{eq:eom-precessional}) numerically when obviously one has a finite $N$. 
In this
case, as shown in Appendix~B, the state of the system is
described by a discrete single-spin density function $P_d({\bf S},t)$,
which to leading order in $N$ may be expanded as
\begin{equation}
P_d({\bf S},t)=P_0({\bf S},t)+\frac{1}{\sqrt{N}}\delta P({\bf S},t).
\label{eq:expansion-1}
\end{equation}
For times $t \ll N$, the time evolution of $P_0$ is given by Eq.~(\ref{eq:Vlasov-1}), 
with that for $\delta P$ given by 
\begin{eqnarray}
&&\frac{\partial \delta P({\bf S},t)}{\partial t} = -\frac{\partial}{\partial{\bf
 S}}\cdot\Big[({\bf S}\times\delta{\bf h}^{{\rm
 eff}})P_0+({\bf S}\times{\bf h}^{{\rm
 eff},0})\delta P\Big],\label{eq:FPE2-again1}
\end{eqnarray}
with $\delta {\bf h}^{\rm eff} \equiv \delta {\bf h}^{\rm eff}[\delta
P]$.
Equivalent to Eq.~(\ref{eq:expansion-1}), one may write 
\begin{equation}
f_d(\theta,\phi,t)=f(\theta,\phi,t)+\frac{1}{\sqrt{N}}\delta
f(\theta,\phi,t),
\label{eq:fd}
\end{equation}
where for times $t \ll N$, one has the time evolution of $f$ given by
Eq.~(\ref{eq:Vlasov-equation}), while as was done in obtaining Eq.~(\ref{eq:Vlasov-equation}), one may show
that the time evolution of $\delta f(\theta,\phi,t) \equiv \delta P({\bf
S},t)$ is obtained from Eq~(\ref{eq:FPE2-again1}) as 
\begin{eqnarray}
&&\fl \frac{\partial \delta f}{\partial t}=
\Big(m_y[\delta f] \cos \phi - m_x[\delta f] \sin\phi
\Big) \frac{\partial f}{\partial \theta} +\Big(m_y[f] \cos \phi - m_x[f] \sin\phi
\Big) \frac{\partial \delta f}{\partial \theta}\nonumber \\
&&
\fl-\Big( m_x[\delta f] \cot\theta \cos\phi+ m_y[\delta f]\cot\theta
\sin\phi - m_z[\delta f] \Big)\frac{\partial
f}{\partial \phi} \nonumber \\
&&
\fl-\Big( m_x[f] \cot\theta \cos\phi+ m_y[f]\cot\theta \sin\phi -
m_z[f]+(2n)D\cos^{2n-1} \theta \Big)\frac{\partial \delta f}{\partial \phi}. 
\label{eq:Vlasov-linearized-again}
\end{eqnarray}

Suppose we choose $f(\theta,\phi,0)$
to be $f_0(\theta,\phi)$ given by Eq.~(\ref{eq:f0}). It then follows from
Eq.~(\ref{eq:Vlasov-equation}) that
$f(\theta,\phi,t)=f(\theta,\phi,0)$, while
Eq.~(\ref{eq:Vlasov-linearized-again}) takes the same form as the
linearized Vlasov equation~(\ref{eq:Vlasov-linearized}):
\begin{eqnarray}
&& \frac{\partial \delta f}{\partial t}=
\Big(m_y[\delta f] \cos \phi - m_x[\delta f] \sin\phi
\Big) \frac{\partial f}{\partial \theta}-(2n)D\cos^{2n-1} \theta \frac{\partial \delta f}{\partial \phi}. 
\label{eq:Vlasov-linearized-again-1}
\end{eqnarray}
Based on our analysis in the preceding
section, we may then conclude that for energies $\epsilon <
\epsilon^\star$, when $f_0(\theta,\phi)$ is an unstable stationary solution
of the Vlasov equation, Eq.~(\ref{eq:fd}) would give
\begin{equation}
(m_x,m_y,m_z)[f_d]=\frac{1}{\sqrt{N}}(m_x,m_y,m_z)[\delta
f].
\end{equation}
Using Eq.~(\ref{eq:m-e<estar-0}) that is a solution of an equation of
the same form, Eq.~(\ref{eq:Vlasov-linearized}), as
Eq.~(\ref{eq:Vlasov-linearized-again-1}), we thus obtain
\begin{equation}
m(t) \sim \frac{1}{\sqrt{N}}e^{\Omega t};~~\epsilon < \epsilon^\star.
\label{eq:m-e<estar}
\end{equation}
Thus, for $\epsilon < \epsilon^\star$, the relaxation time 
over which the magnetization acquires a value of
$O(1)$ scales as $\log N$. On the other hand, for energies $\epsilon >
\epsilon^\star$, when $f_0(\theta,\phi)$ is Vlasov-stationary and
stable, the system would remain unmagnetized for times $t\ll N$. 
In this case, it is known that for longer times, the time evolution is described by~(see Appendix~B):
\begin{eqnarray} 
&& 
\frac{\partial P_0({\bf S},t)}{\partial
t} +\frac{\partial}{\partial{\bf
 S}}\cdot({\bf S}\times{\bf h}^{{\rm
 eff},0})P_0=-\frac{1}{N}\Big\langle\frac{\partial}{\partial{\bf
 S}}\cdot({\bf S}\times\delta{\bf h}^{{\rm
 eff}})\delta P\Big\rangle.\label{eq:FPE-final-LB-0}
\end{eqnarray}
Then, for $\epsilon > \epsilon^\star$, only
for longer times of order $N$ when the dynamics~(\ref{eq:FPE-final-LB-0}) comes into play would there be an
evolution of the initial unmagnetized state. Consequently, the state
$f_0(\theta,\phi)$ manifests itself in a finite system as a long-lived
QSS that evolves very slowly, that is, over a timescale that diverges
with $N$. 

\subsection{Numerical results}
\label{sec:numerics}

Here, we discuss numerical results in support of our theoretical
analysis of the preceding section. We present our results for two representative
values of $n$, namely, $n=1,2$. In performing numerical integration of
the dynamics~(\ref{eq:eom-precessional}), unless stated otherwise, we
employ a fourth-order Runge-Kutta integration
algorithm with timestep equal to $0.01$. In the numerical results that
we present, data averaging has been typically over several hundreds to
thousand runs of the dynamics starting from different realizations of
the FD state~(\ref{eq:FD}). 

\begin{figure}[!ht]
\centering
\includegraphics[width=7cm]{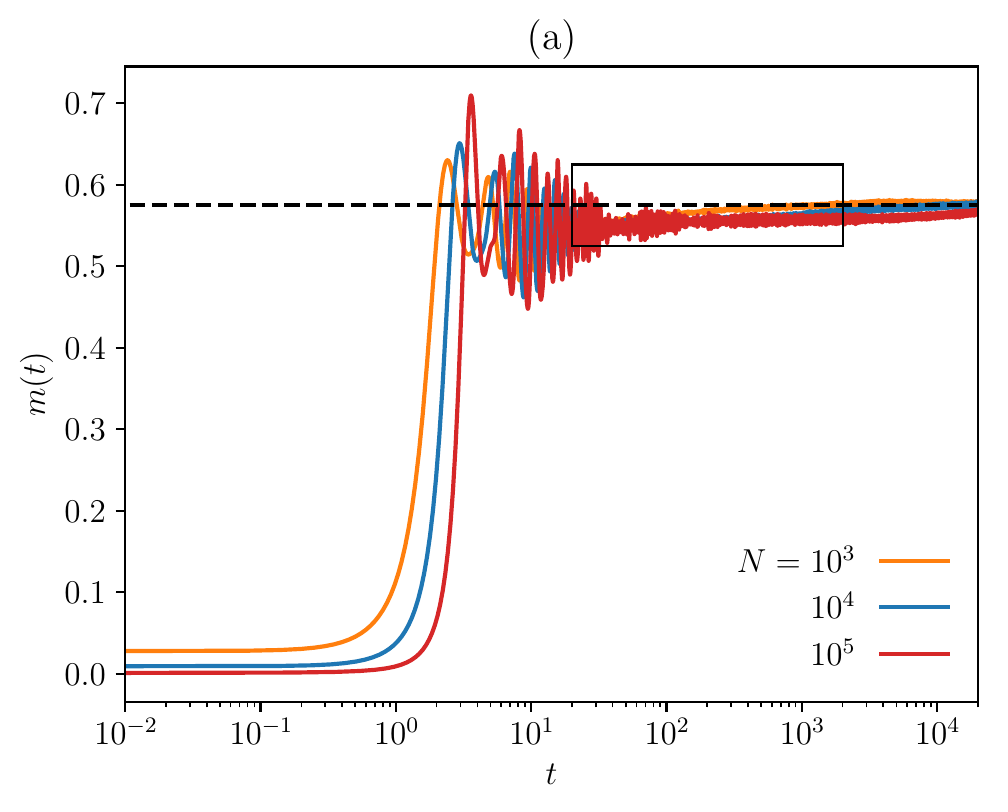}
        \includegraphics[width=7cm]{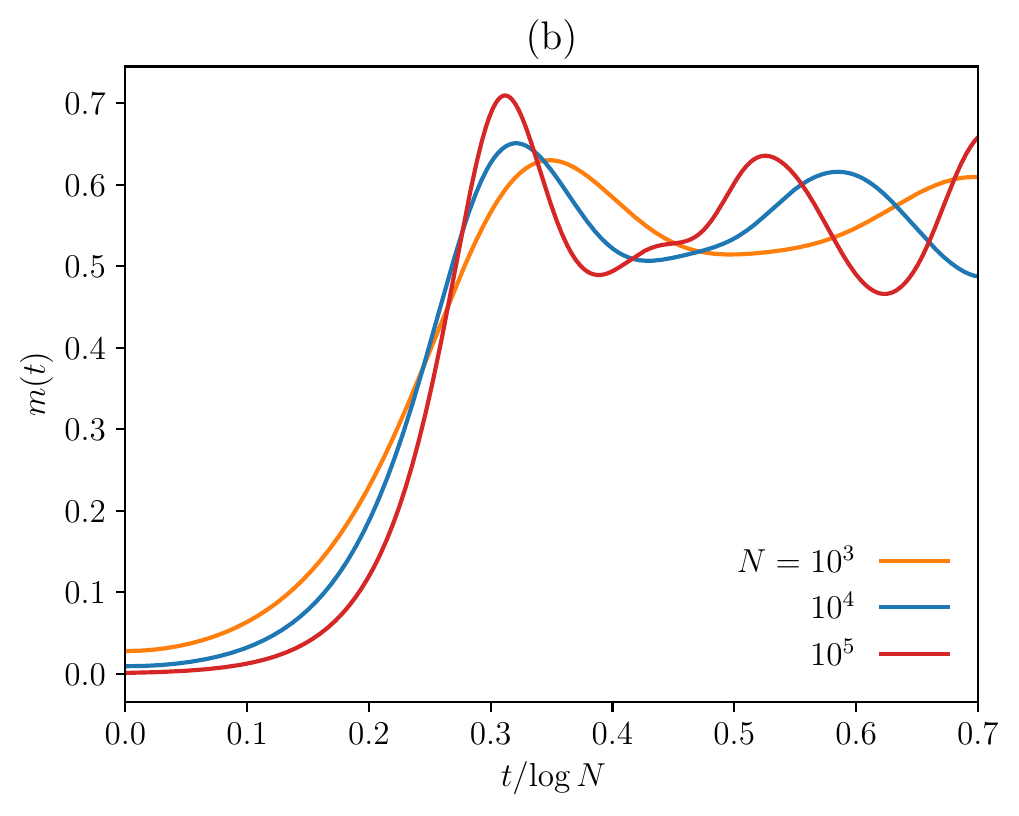}
        \includegraphics[width=7cm]{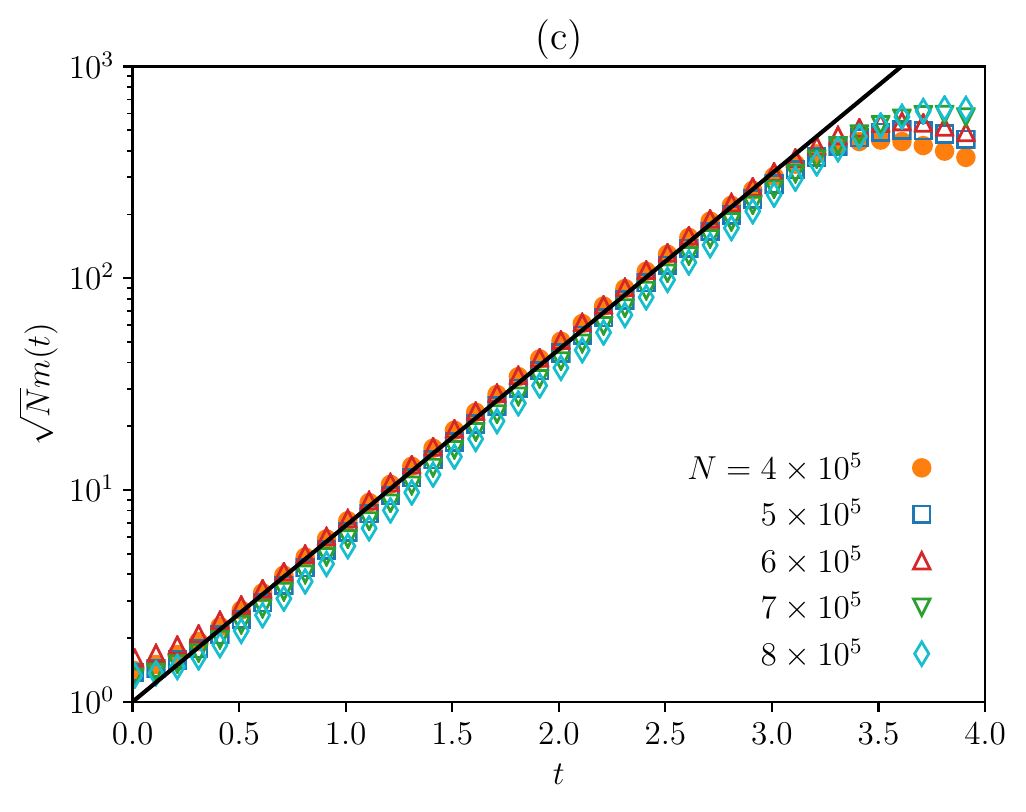}
\caption{For the model~(\ref{eq:H}) with $n=1$ and $D=5.0$ thus yielding
$\epsilon_c\approx 0.2381$, the figure shows the relaxation under the
deterministic dynamics~(\ref{eq:eom-precessional}) of an initial
nonmagnetized FD
state~(\ref{eq:FD}) with $\beta_{\rm FD}=1000$ (thus yielding stability threshold
$\epsilon^\star \approx 0.0795$) for energy $\epsilon < \epsilon^\star$; here, we have chosen
$\epsilon=0.0212$. One may observe a fast relaxation out of the initial
FD state (panel (a)) over a time that scales with $N$ as $\log N$ (panel (b)). In
panel (a), the dashed line represents the value of equilibrium
magnetization at the studied energy value. The
initial fast growth of the magnetization observed in (a) follows
Eq.~(\ref{eq:m-e<estar}), as is evident from the data collapse for the
scaled magnetization $\sqrt{N}m(t)$ as a function of $t$ shown in panel
(c). Here, the black line
represents $e^{\Omega t}$, with $\Omega$ obtained as the magnitude of
imaginary part of the root of Eq.~(\ref{eq:n2-omega-eqn1}) for which the
imaginary part is the largest in magnitude; Here, we have $\Omega
\approx 1.915$.}
\label{fig:s2-fig1-nonoise}
\end{figure}

\begin{figure}
\centering
\includegraphics[width=7cm]{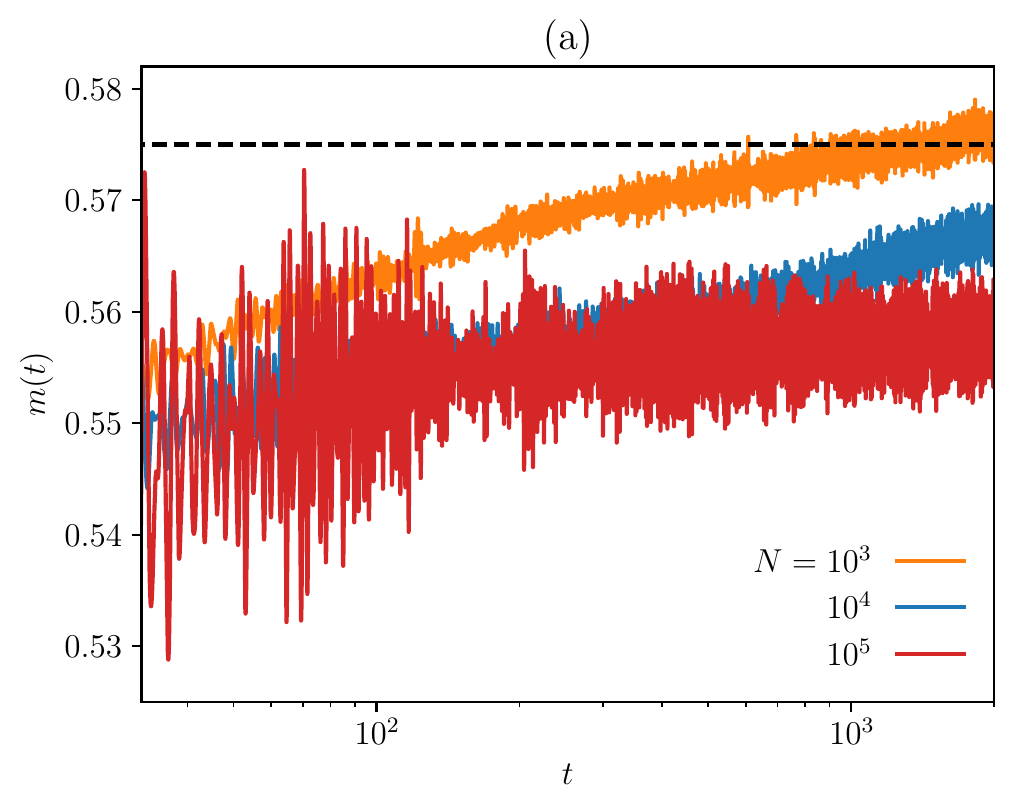}
\includegraphics[width=7cm]{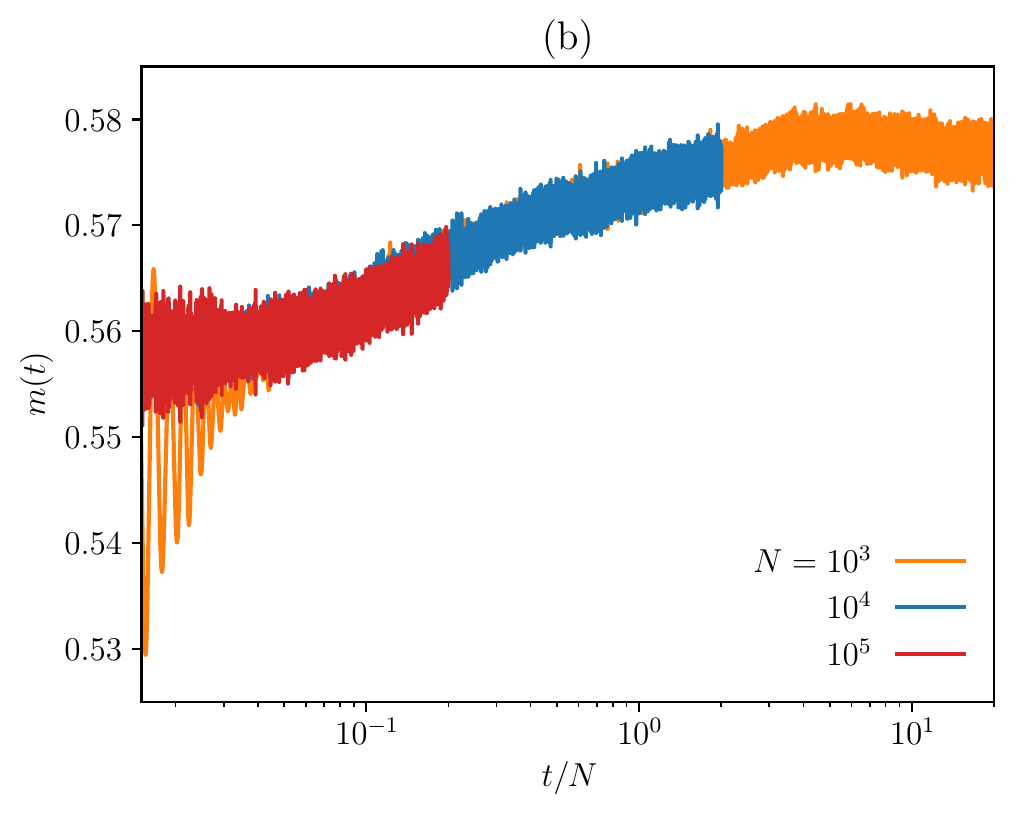}
\caption{For the same parameter values as in
Fig.~\ref{fig:s2-fig1-nonoise}, panel
(a) shows for the case of relaxation under deterministic
evolution~(\ref{eq:eom-precessional}) of the nonmagnetized FD state~(\ref{eq:FD}) a zoom in on 
the boxed part of Fig.~\ref{fig:s2-fig1-nonoise}(a). Here, the dashed line
represents the value of equilibrium magnetization at the studied energy
value. The plot suggests the existence of a magnetized QSS with a
lifetime that scales linearly with $N$ (see panel (b)).}
\label{fig:s2-fig2-nonoise}
\end{figure}

We first discuss the results for $n=1$, for which we make the choice
$D=5.0$ that yields the equilibrium critical energy $\epsilon_c \approx
0.2381$. Choosing as an initial condition the nonmagnetized FD
state~(\ref{eq:FD}) with $\beta=1000$, for which one has the stability
threshold $\epsilon^\star \approx 0.0795$ (see
Fig.~\ref{fig:phase-diagram}), Fig.~\ref{fig:s2-fig1-nonoise}(a) shows for
energy $\epsilon < \epsilon^\star$ a fast relaxation out of the initial
state on a timescale $\sim \log N$ (see Fig.~\ref{fig:s2-fig1-nonoise}(b)). This is consistent with the
prediction based on Eq.~(\ref{eq:m-e<estar}), which is further validated
by the collapse of the data for $\sqrt{N} m(t)$ vs. $t$ for different
values of $N$ shown in
Fig.~\ref{fig:s2-fig1-nonoise}(c); here, the growth
rate $\Omega$ of $m(t)$ is obtained as the magnitude of imaginary part of the root of Eq.~(\ref{eq:n2-omega-eqn1}) for which the imaginary part is the largest in magnitude.
Figure~\ref{fig:s2-fig2-nonoise}(a) shows that the
relaxation observed in Fig.~\ref{fig:s2-fig1-nonoise} out of the initial FD
state is not to Boltzmann-Gibbs equilibrium but is to a magnetized QSS that has a
lifetime that scales linearly with $N$, Fig.~\ref{fig:s2-fig2-nonoise}(b).
Summarizing, for energy $\epsilon< \epsilon^\star$, relaxation of
nonmagnetized FD state to Boltzmann-Gibbs equilibrium is a two-step process: in the
first step, the system relaxes over a timescale $\sim \log N$ to a magnetized
QSS, while in the second step, this QSS relaxes over a timescale $\sim
N$ to Boltzmann-Gibbs equilibrium.

\begin{figure}[!ht]
\centering
        \includegraphics[width=7cm]{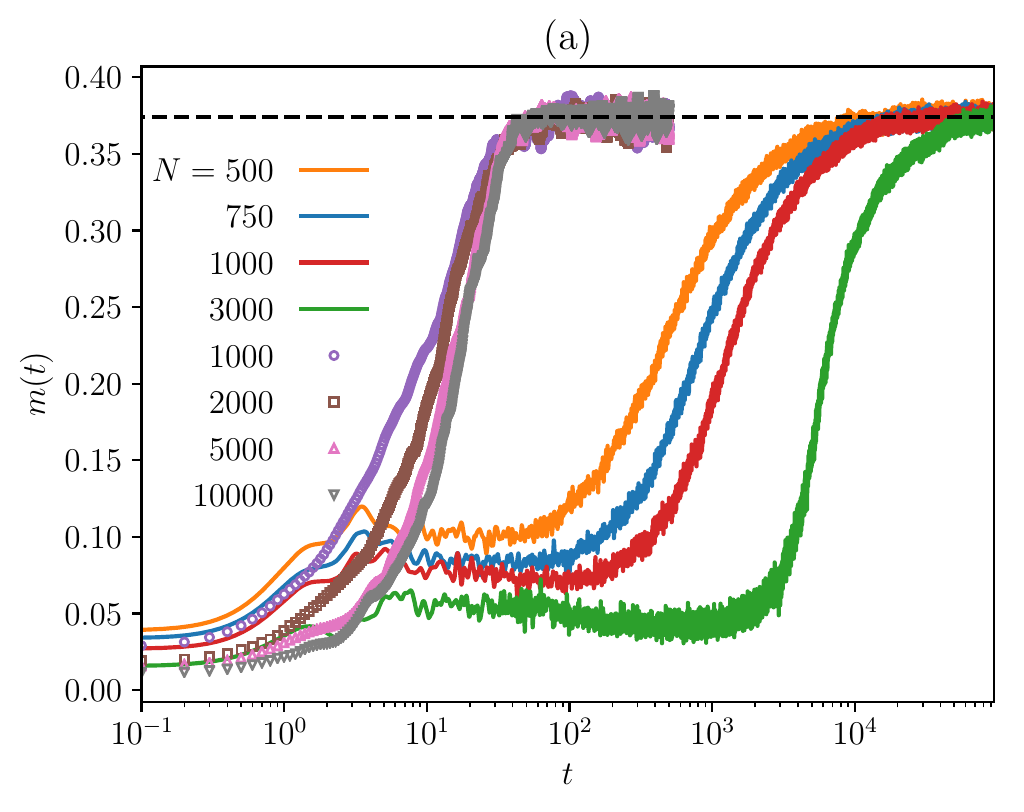}
        \includegraphics[width=7cm]{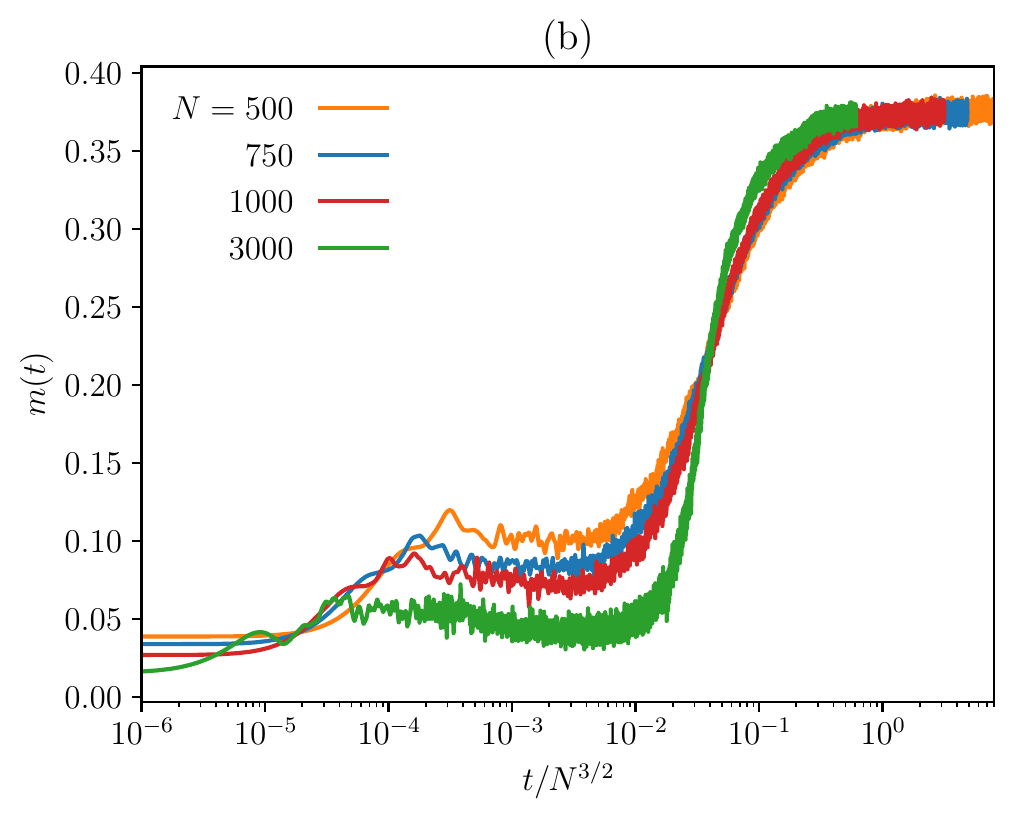}
\caption{For the model~(\ref{eq:H}) with $n=1$ and $D=5.0$ thus yielding
$\epsilon_c\approx 0.2381$, the figure shows in lines the relaxation under the
deterministic dynamics~(\ref{eq:eom-precessional}) of an initial
nonmagnetized FD
state~(\ref{eq:FD}) with $\beta_{\rm FD}=1000$ (thus yielding stability threshold
$\epsilon^\star \approx 0.0795$) for energy $\epsilon^\star < \epsilon <
\epsilon_c$; here, we have chosen
$\epsilon=0.1473$. Here, the dashed line
represents the value of equilibrium magnetization at the studied energy
value. One may observe the existence of a nonmagnetized QSS
with a lifetime that diverges with the system size as $N^{3/2}$ (panel
(b)). The points in panel (a) denote results based on numerical integration of the stochastic
dynamics~(\ref{eq:eom-noise}) for $\gamma=0.05$ and at a value of
temperature such that one obtains the same value of the
equilibrium magnetization as that obtained at the value of energy chosen
for the deterministic dynamics studied in (a). The results imply a
fast relaxation to equilibrium on a timescale that does not depend on
$N$.}
\label{fig:s2-fig3-nonoise}
\end{figure}

\begin{figure}[!ht]
\centering
        \includegraphics[width=7cm]{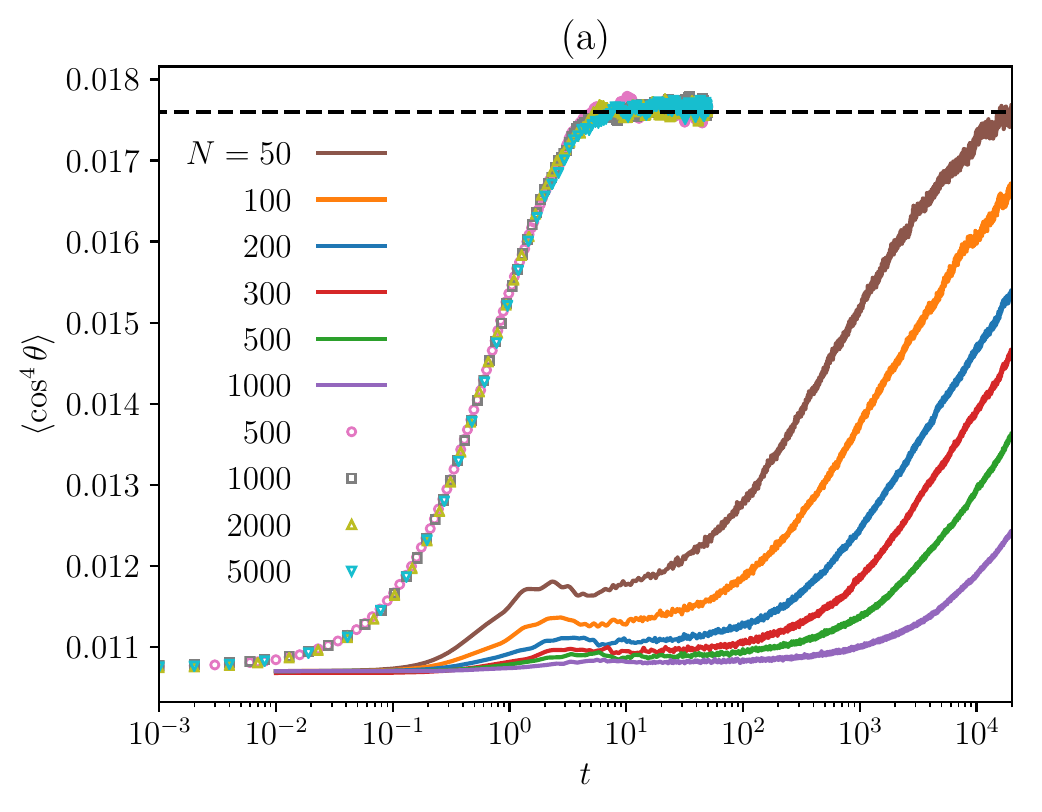}
        \includegraphics[width=7cm]{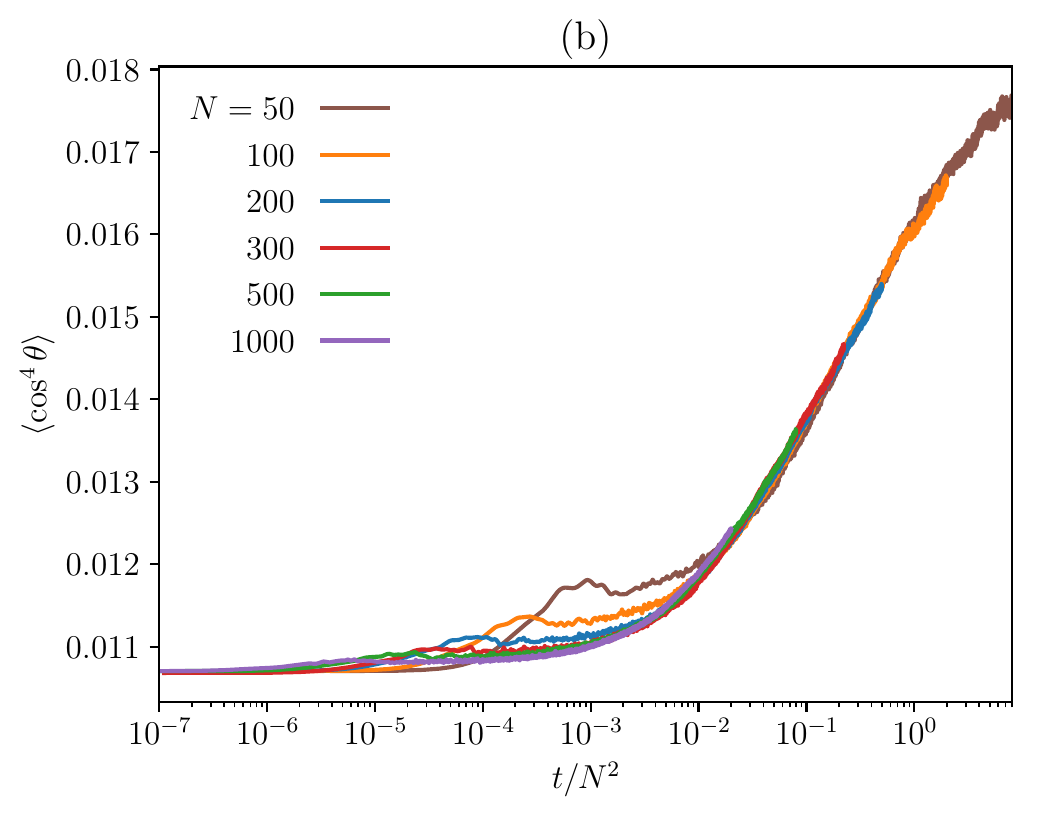}
\caption{For the model~(\ref{eq:H}) with $n=1$ and $D=5.0$ thus yielding
$\epsilon_c\approx 0.2381$, the figure shows in lines the relaxation under the
deterministic dynamics~(\ref{eq:eom-precessional}) of an initial
nonmagnetized FD
state~(\ref{eq:FD}) with $\beta_{\rm FD}=1000$ (thus yielding stability threshold
$\epsilon^\star \approx 0.0795$) for energy $\epsilon > \epsilon_c$; here, we have chosen
$\epsilon=0.3863$. Here the dashed line denotes the equilibrium value of
$\langle \cos^4\theta\rangle$ at the studied energy value, which may be
obtained from the analysis in Section~\ref{sec:equilibrium-properties}. One may observe the existence of a nonmagnetized QSS
with a lifetime that diverges with the system size as $N^{2}$ (panel
(b)). The points in panel (a) represent results obtained from numerical integration of the stochastic
dynamics~(\ref{eq:eom-noise}) for $\gamma=0.05$ and at a value of
temperature for which one obtains the same value of equilibrium $\langle \cos^4\theta\rangle$ as that obtained at the value of energy chosen
for the deterministic dynamics studied in (a). From the results, one may
conclude a fast relaxation to equilibrium on a size-independent timescale, with no
sign of quasistationarity.}
\label{fig:s2-fig4-nonoise}
\end{figure}

For energies $\epsilon^\star < \epsilon < \epsilon_c$,
Fig.~\ref{fig:s2-fig3-nonoise}(a) shows that consistent with our analysis, the initial FD state appears as
a nonmagnetized QSS that relaxes to Boltzmann-Gibbs equilibrium over a time which by
virtue of the data presented in Fig.~\ref{fig:s2-fig3-nonoise}(b) may be
concluded to be scaling with $N$ as $N^{3/2}$. For energies $\epsilon >
\epsilon_c$ too is the initial FD state a stable stationary solution of
the Vlasov equation, and is expected to show up as a QSS. However, here
the magnetization is not the right quantity to monitor since both the FD
state and Boltzmann-Gibbs equilibrium are nonmagnetized. Consequently, we choose
$\langle \cos^4 \theta\rangle = (1/N)\sum_{i=1}^N \cos^4 \theta_i$ to
monitor as a
function of time (note that for $\epsilon > \epsilon_c$, the quantity
$\langle \cos^2\theta\rangle$ is strictly a constant for infinite $N$,
showing fluctuations about this constant value for finite $N$).
Figure~\ref{fig:s2-fig4-nonoise}
shows that indeed the initial FD state does show up as a QSS that has a
lifetime that scales quadratically with $N$. In all cases reported above and in the following when we observe an
initial QSS with zero magnetization relaxing eventually to a magnetized
state in equilibrium, it may be noted that due to the symmetry of the
Hamiltonian~(\ref{eq:H}) under spin rotation about the $z$-axis, the particular direction the
magnetization chooses in equilibrium may depend on the particular
realization of the QSS under study. The equilibrium magnetization vector may
even have some rotation in time, and only the application of an external
field may select a given orientation of the vector.

\begin{figure}[!ht]
\centering
        \includegraphics[width=7cm]{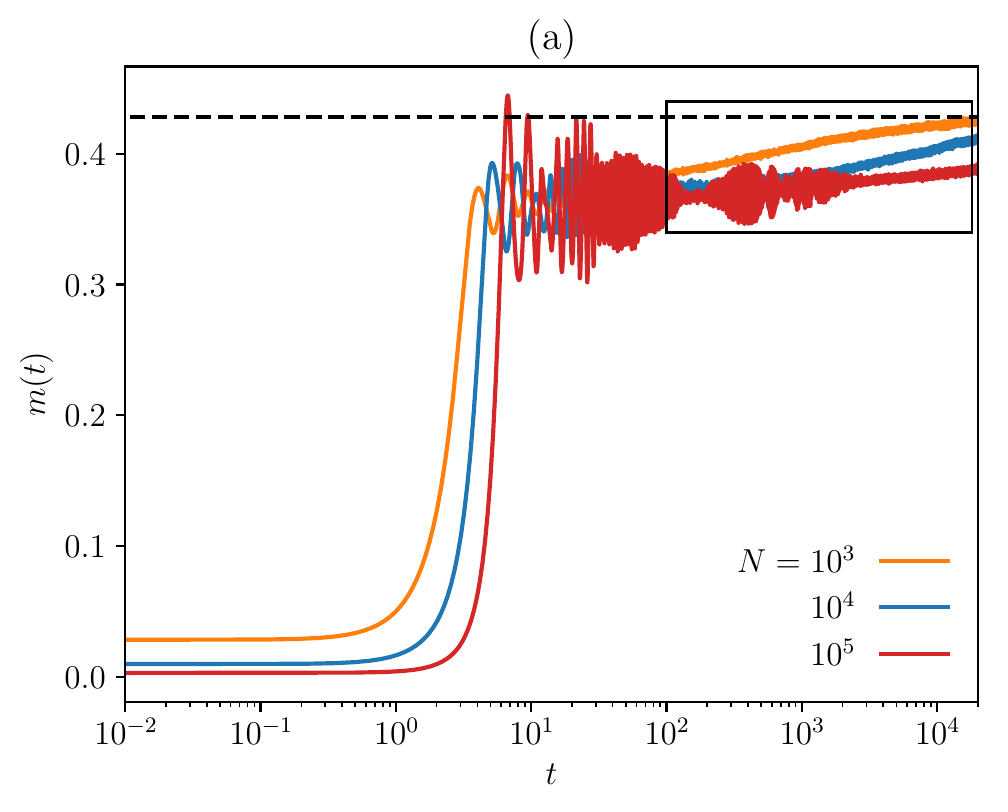}
        \includegraphics[width=7cm]{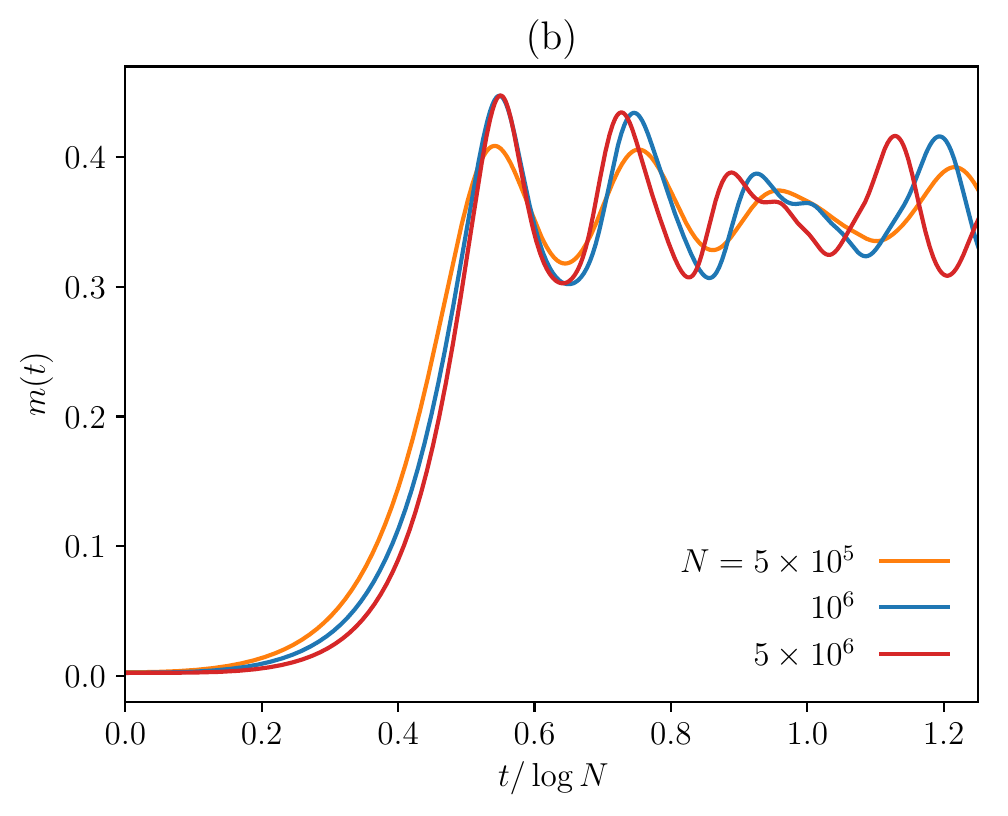}
        \includegraphics[width=7cm]{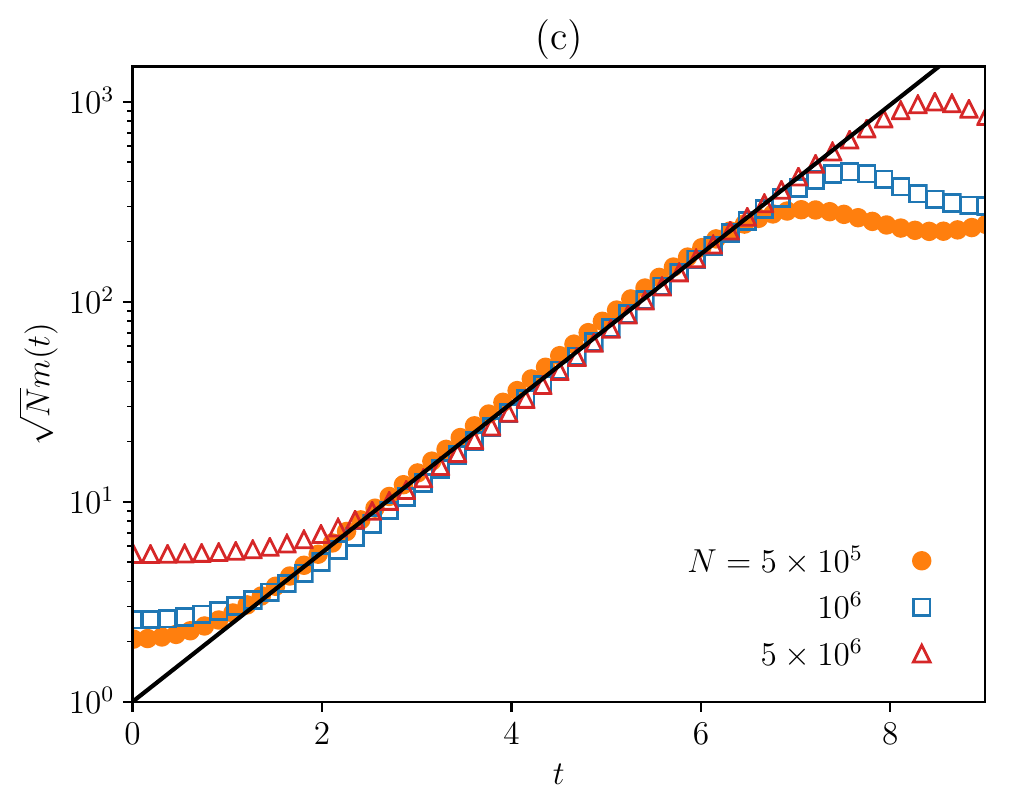}
\caption{For the model~(\ref{eq:H}) with $n=2$ and $D=15.0$ thus yielding
$\epsilon_c\approx 0.1175$, the figure shows the relaxation under the
deterministic dynamics~(\ref{eq:eom-precessional}) of an initial
nonmagnetized FD
state~(\ref{eq:FD}) with $\beta_{\rm FD}=100$ (thus yielding stability threshold
$\epsilon^\star \approx 0.0277$) for energy $\epsilon < \epsilon^\star$; here, we have chosen
$\epsilon=0.0111$. One may observe a fast relaxation out of the initial
FD state (panel (a)) over a time that scales with $N$ as $\log N$ (panel (b)). In
panel (a), the dashed line represents the value of equilibrium
magnetization at the studied energy value. The
initial fast growth of the magnetization observed in (a) follows
Eq.~(\ref{eq:m-e<estar}), as is evident from the data collapse for the
scaled magnetization $\sqrt{N}m(t)$ as a function of $t$ shown in panel
(c). Here, the black line
represents $e^{\Omega t}$, with $\Omega$ obtained as the magnitude of
imaginary part of the root of Eq.~(\ref{eq:n4-omega-eqn}) for which the
imaginary part is the largest in magnitude; Here, we have $\Omega
\approx 0.859$.}
\label{fig:s4-fig1-nonoise}
\end{figure}

\begin{figure}[!ht]
\centering
        \includegraphics[width=7cm]{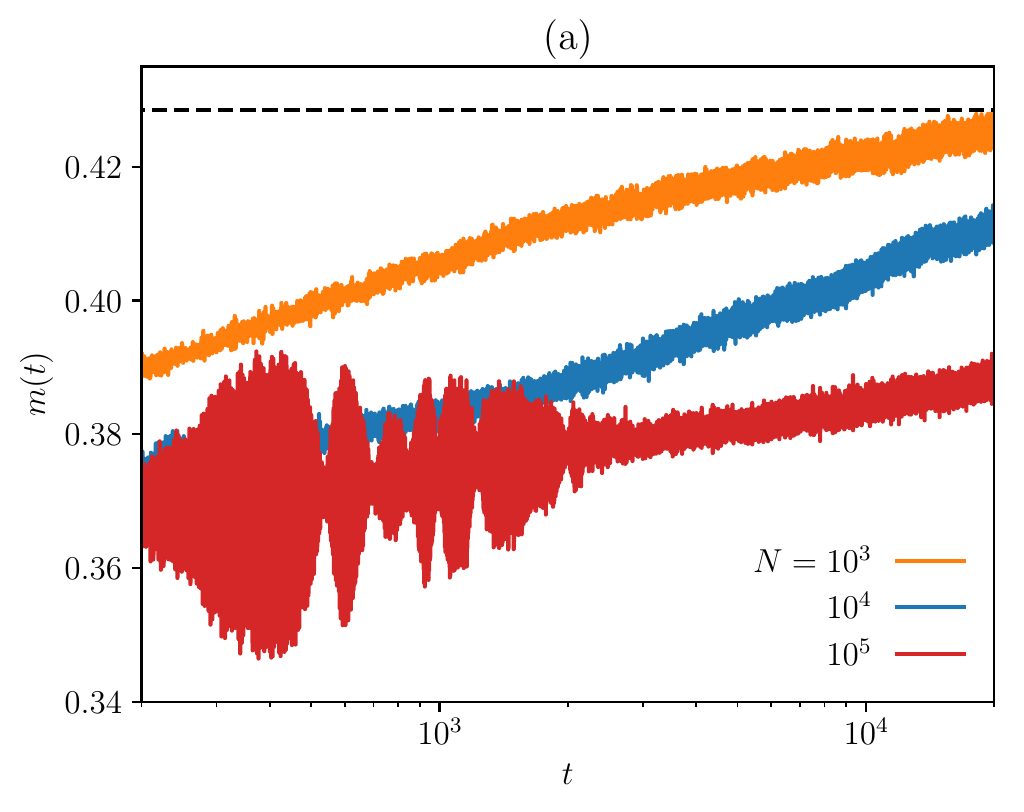}
        \includegraphics[width=7cm]{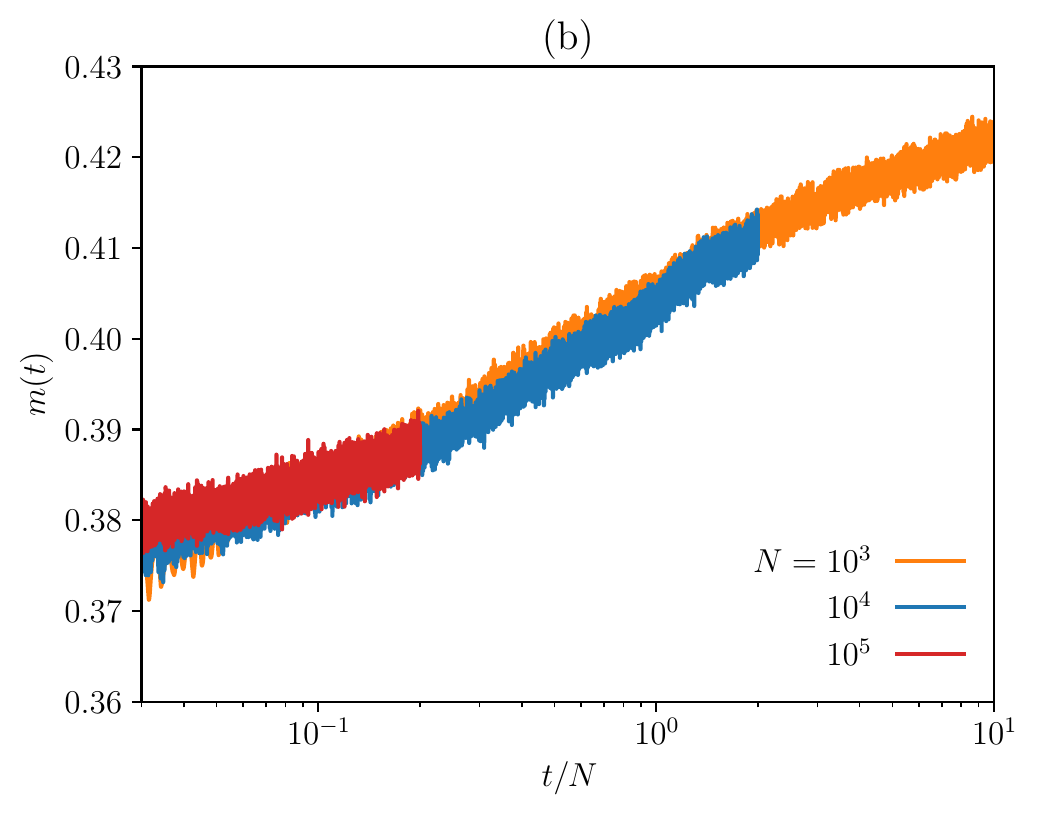}
\caption{For the same parameter values as in
Fig.~\ref{fig:s4-fig1-nonoise}, panel
(a) shows for the case of relaxation under deterministic
evolution~(\ref{eq:eom-precessional}) of the nonmagnetized FD state~(\ref{eq:FD}) a zoom in on 
the boxed part of Fig.~\ref{fig:s4-fig1-nonoise}(a). Here, the dashed line
represents the value of equilibrium magnetization at the studied energy
value. The plot suggests the existence of a magnetized QSS with a
lifetime that scales linearly with $N$ (see panel (b)).}
\label{fig:s4-fig2-nonoise}
\end{figure}

\begin{figure}[!ht]
\centering
        \includegraphics[width=7cm]{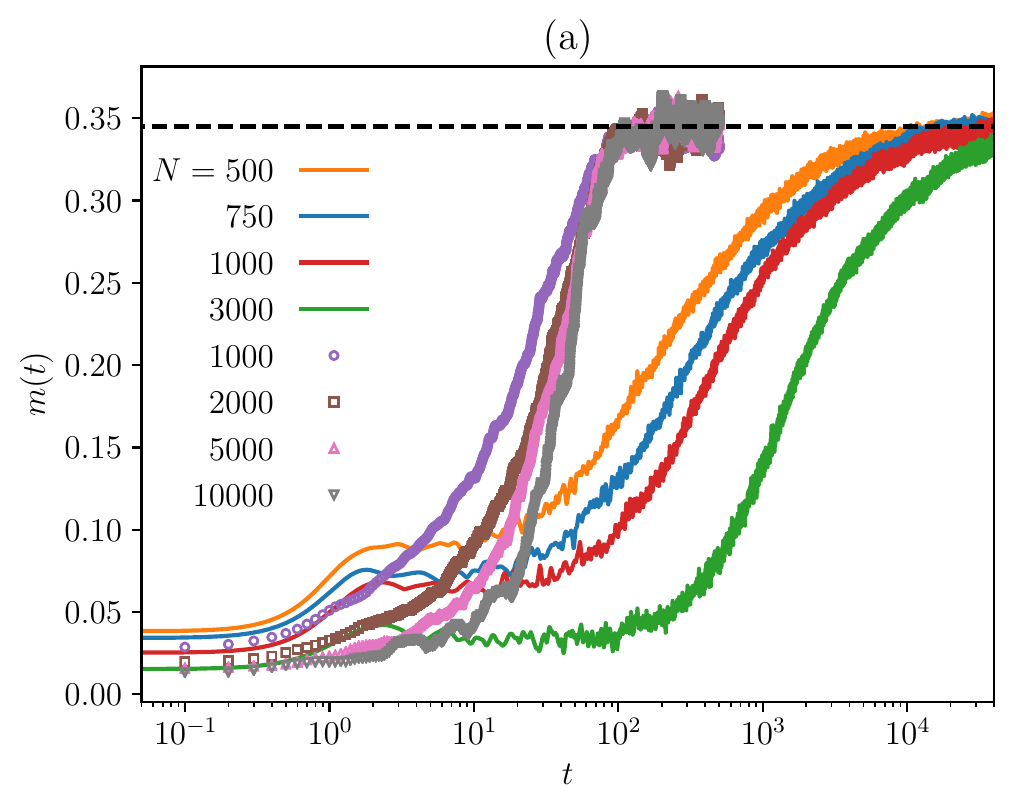}
        \includegraphics[width=7cm]{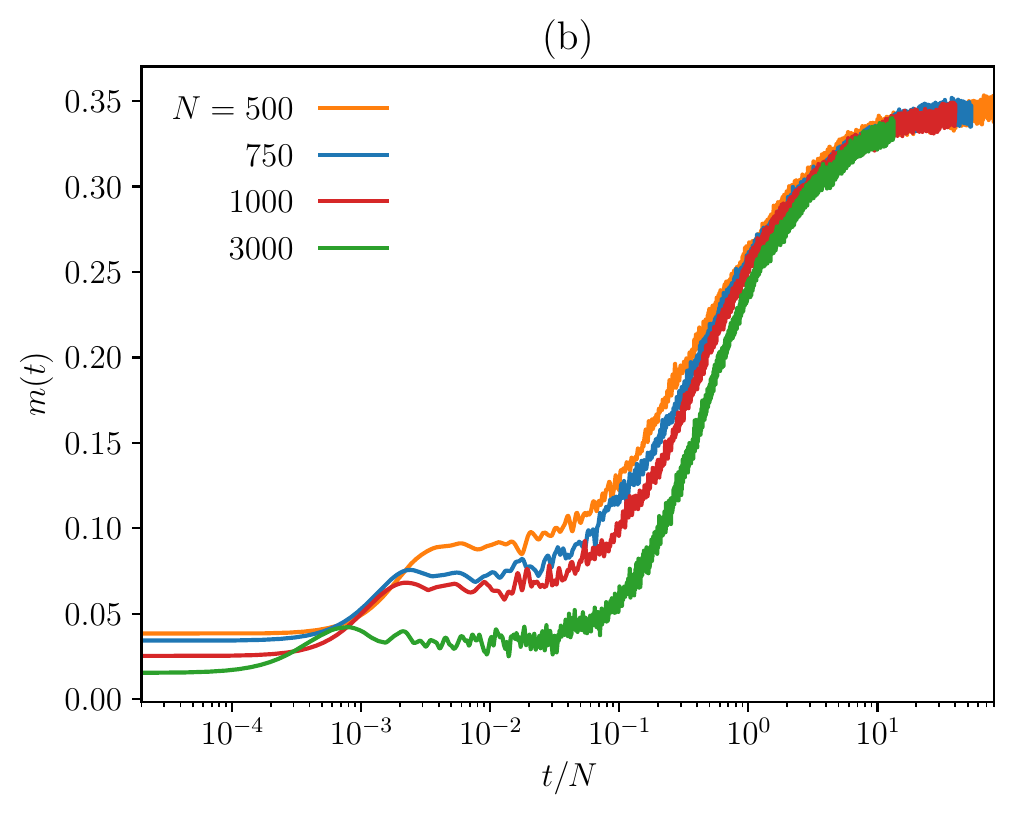}
        \caption{For the model~(\ref{eq:H}) with $n=2$ and $D=15.0$ thus yielding
$\epsilon_c\approx 0.1175$, the figure shows in lines the relaxation under the
deterministic dynamics~(\ref{eq:eom-precessional}) of an initial
nonmagnetized FD
state~(\ref{eq:FD}) with $\beta_{\rm FD}=100$ (thus yielding stability threshold
$\epsilon^\star \approx 0.0277$) for energy $\epsilon^\star < \epsilon <
\epsilon_c$; here, we have chosen
$\epsilon=0.0487$. Here, the dashed line
represents the value of equilibrium magnetization at the studied energy
value. One may observe the existence of a nonmagnetized QSS
with a lifetime that diverges with the system size as $N$ (panel
(b)). The points in panel (a) denote results based on numerical integration of the stochastic
dynamics~(\ref{eq:eom-noise}) for $\gamma=0.05$ and at a value of
temperature such that one obtains the same value of the
equilibrium magnetization as that obtained at the value of energy chosen
for the deterministic dynamics studied in (a). The results imply a
fast relaxation to equilibrium on a timescale that does not depend on
$N$.}
\label{fig:s4-fig3-nonoise}
\end{figure}

\begin{figure}[!ht]
\centering
        \includegraphics[width=7cm]{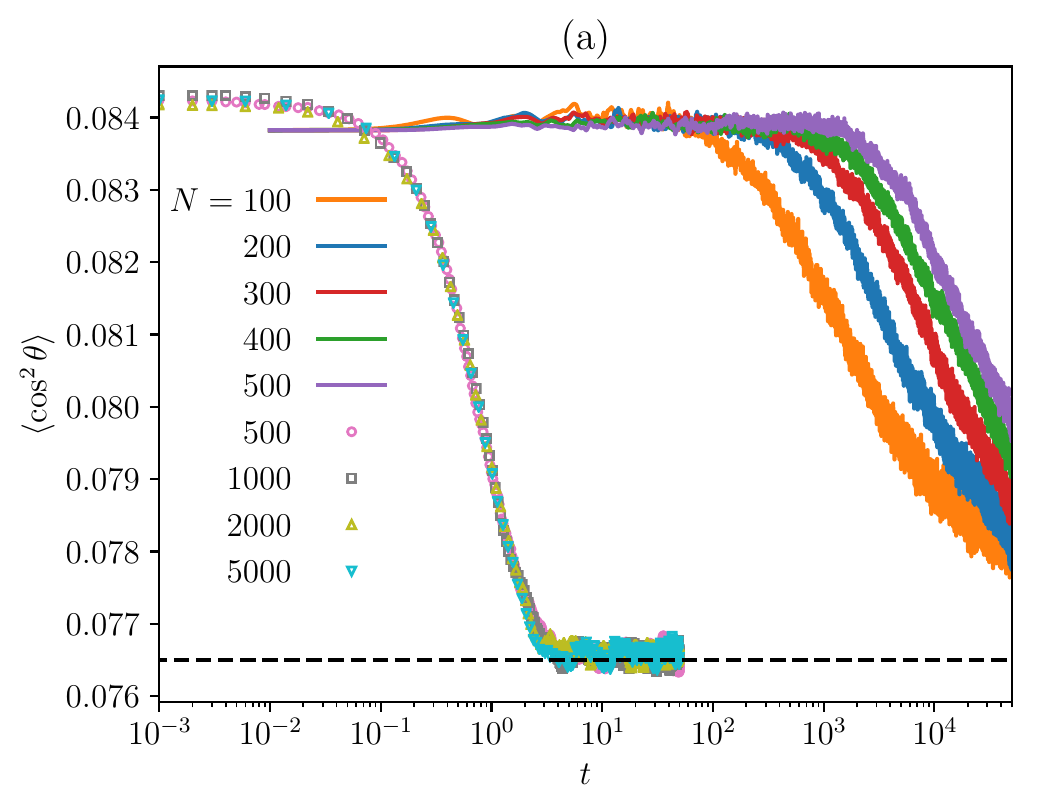}
        \includegraphics[width=7cm]{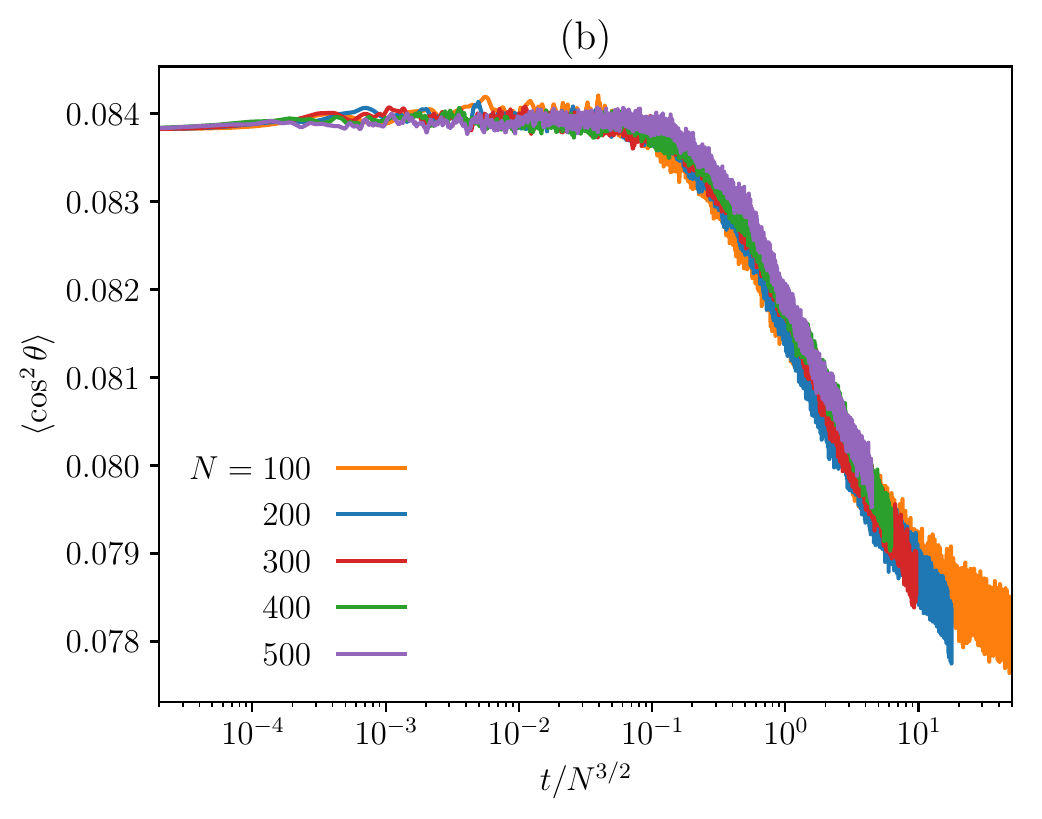}
\caption{For the model~(\ref{eq:H}) with $n=2$ and $D=15.0$ thus yielding
$\epsilon_c\approx 0.1175$, the figure shows in lines the relaxation under the
deterministic dynamics~(\ref{eq:eom-precessional}) of an initial
nonmagnetized FD
state~(\ref{eq:FD}) with $\beta_{\rm FD}=100$ (thus yielding stability threshold
$\epsilon^\star=0.0277$) for energy $\epsilon > \epsilon_c$; here, we have chosen
$\epsilon=0.1925$. Here the dashed line denotes the equilibrium value of
$\langle \cos^2\theta\rangle$ at the studied energy value, which may be
obtained from the analysis in Section~\ref{sec:equilibrium-properties}. One may observe the existence of a nonmagnetized QSS
with a lifetime that diverges with the system size as $N^{3/2}$ (panel
(b)). The points in panel (a) represent results obtained from numerical integration of the stochastic
dynamics~(\ref{eq:eom-noise}) for $\gamma=0.05$ and at a value of
temperature for which one obtains the same value of 
equilibrium $\langle \cos^2\theta \rangle$ as that obtained at the value of energy chosen
for the deterministic dynamics studied in (a). From the results, one may
conclude a fast relaxation to equilibrium on a size-independent timescale, with no
sign of quasistationarity.}
\label{fig:s4-fig4-nonoise}
\end{figure}

To demonstrate that the aforementioned relaxation scenario is quite
generic to the model~(\ref{eq:H}), we now present in
Figs.~\ref{fig:s4-fig1-nonoise}
--~\ref{fig:s4-fig4-nonoise} results for another
value of $n$, namely, $n=2$. As may be observed from the figures, one
has the same qualitative features of the relaxation process as that
discussed above for $n=1$. Note that for energies $\epsilon >
\epsilon_c$, one has in contrast to the $n=1$ case the quantity $\langle
\cos^4 \theta \rangle$ a constant in time and consequently one monitors
$\langle \cos^2\theta \rangle$ as a function of time, see
Fig.~\ref{fig:s4-fig4-nonoise}. Differences from the $n=1$ case appear in specific scalings of
QSSs: the nonmagnetized QSS occurring for energies $\epsilon^\star <
\epsilon < \epsilon_c$ has a lifetime scaling as $N$, while the one
occurring for energies $\epsilon > \epsilon_c$ has a lifetime growing
with $N$ as $N^{3/2}$.

\section{Analysis of the stochastic dynamics~(\ref{eq:eom-noise})}
\label{sec:noisy}

\subsection{Behavior in the limit $N \to \infty$}
\label{sec:noisy-infiniteN}

The stochastic dynamics~(\ref{eq:eom-noise}) in the limit
$N \to \infty$ may be studied by considering the time evolution of the
single-spin distribution function $P_0({\bf
S},t)$ derived in Appendix~B as
\begin{eqnarray}
&& 
\fl \frac{\partial P_0({\bf S},t)}{\partial
t} +\frac{\partial}{\partial{\bf
 S}}\cdot({\bf S}\times{\bf h}^{{\rm
 eff},0})P_0 = \gamma\frac{\partial}{\partial{\bf
 S}}\cdot\Big[({\bf
 S}\times{\bf S}\times{\bf h}^{{\rm eff},0})-(1/\beta)({\bf
 S}\times({\bf S}\times\frac{\partial}{\partial{\bf
 S}}))\Big]P_0. \nonumber \\
 \label{eq:time-evolution-stochastic}
\end{eqnarray}
Note that the state~(\ref{eq:FD}), or more generally, the state
(\ref{eq:f0}), is
not a stationary solution of Eq.~(\ref{eq:time-evolution-stochastic}),
while, as already discussed, they both solve the energy-conserving Vlasov
dynamics~(\ref{eq:Vlasov-1}) in the stationary state. From the structure
of the above equation, it follows that for times $t \ll 1/\gamma$, one may neglect the right
hand side, and consequently, the time evolution is governed solely by the
left hand side set to zero, which is nothing but the Vlasov
equation~(\ref{eq:Vlasov-1}). As a result, the energy is conserved for
times $t \ll 1/\gamma$, and, based on the analysis
presented in Section~\ref{sec:noiseless-infiniteN}, the state~(\ref{eq:FD}) appears as an unstable
stationary state for energies $\epsilon < \epsilon^\star$ and as a stable stationary state for energies $\epsilon > \epsilon^\star$. For times of order $1/\gamma$, we may however not
neglect the right hand side of Eq.~(\ref{eq:time-evolution-stochastic}),
and hence, we would observe the energy to be changing over times of
$O(1/\gamma)$ and the state~(\ref{eq:FD}) to be evolving to relax to the
stationary state of Eq.~(\ref{eq:time-evolution-stochastic}), which is
nothing but the Boltzmann-Gibbs equilibrium, see Appendix~B. 
\subsection{Behavior for finite $N$}
\label{sec:noisy-finiteN}
In this case, finite-$N$ corrections need to be added to
Eq.~(\ref{eq:time-evolution-stochastic}), and as shown in Appendix~B, the time
evolution is instead given by
\begin{eqnarray}
&& 
\fl \frac{\partial P_0({\bf S},t)}{\partial
t} +\frac{\partial}{\partial{\bf
 S}}\cdot({\bf S}\times{\bf h}^{{\rm
 eff},0})P_0 = \gamma\frac{\partial}{\partial{\bf
 S}}\cdot\Big[({\bf
 S}\times{\bf S}\times{\bf h}^{{\rm eff},0})-(1/\beta)({\bf
 S}\times({\bf S}\times\frac{\partial}{\partial{\bf
 S}}))\Big]P_0\nonumber \\
 &&\fl-\frac{1}{N}\Big\langle\frac{\partial}{\partial{\bf
 S}}\cdot\Big[({\bf S}\times\delta{\bf h}^{{\rm
 eff}})\delta P-\gamma({\bf
 S}\times{\bf S}\times\delta{\bf h}^{{\rm eff}})\delta
 P\Big]\Big\rangle.\label{eq:FPE-final-1}
\end{eqnarray}
Then, based on our previous analysis, we may conclude that for a given
$N$, when the noise is strong enough
that $1/\gamma \ll 
N$, the dynamics~(\ref{eq:FPE-final-1}) would be dominated by the first
term on the right hand side. As a result, over times $t \sim 1/\gamma$, the state~(\ref{eq:FD}) would
relax to the Boltzmann-Gibbs equilibrium state, and no size-dependent
relaxation and hence QSSs should be
expected. What happens in the opposite limit, that is, for $1/\gamma
\gg N$? Then, over
times of $O(N)$, the QSS observed for times $t\ll N$, would start
evolving towards Boltzmann-Gibbs equilibrium. The relaxation would be
further assisted by the effects of noise that come into effect over
times of order $1/\gamma$. On the basis of the foregoing, we may expect
that for a given $N$, as one tunes $\gamma$ from very small to very
large values, one should see a cross-over behavior, from a
size-dependent relaxation at small $\gamma$ to a size-independent one at
large $\gamma$.

\subsection{An alternative to dynamics~(\ref{eq:eom-noise}): A Monte
Carlo dynamical scheme}
\label{sec:MC}
An alternative way of modeling the effect of environment-induced noise
on the dynamics~(\ref{eq:eom}) is to invoke a Monte Carlo update scheme of the spin
values that guarantees that the long-time state of the system is
Boltzmann-Gibbs 
equilibrium. In
this scheme, randomly selected spins attempt to rotate by a stipulated
amount (which itself could be random) with a probability that depends on
the change in the energy of the system as a result of the attempted
update of the state of the system~\cite{Glauber:1963,Krapivsky:2010}. 
Specifically, to perform the Monte Carlo dynamics at temperature
$T=1/\beta$, one implements the following steps~\cite{Slanic:1991}:
\begin{enumerate}
\item One starts with a spin configuration in the nonmagnetized FD
state.
\item Next, one
selects a spin at random and attempts to change its direction at
random, that is, choose a value of $\theta$ uniformly in $[0,\pi]$ and a
value of $\phi$ uniformly in $[0,2\pi)$ and assign these values to the
spin.
\item One then computes $\Delta E$, the change in the energy of the system that this
attempted change of spin direction results in.
\item If $\Delta E<0$, that is, the system energy is lowered by the
change of spin direction, the change is accepted. 
\item On the other hand, if the energy increases by changing the spin
direction, that is, $\Delta E>0$, one computes the Boltzmann probability
$p =
\exp( - \beta \Delta E)$. Next, if a random number $r$ chosen uniformly in
$[0,1]$ satisfies $r < p$, the change in spin direction is accepted;
otherwise, the attempted change is rejected and the previous spin configuration is retained.
\item Time is measured in units of Monte Carlo steps (MCS), where one
step corresponds to $N$ attempted changes in spin direction.
\item At the end of every MCS, one computes the desired physical quantities such as the magnetization.
In practice, one repeats steps (ii) -- (v) to obtain values as a function of
time of these physical quantities averaged over a sufficient number
of independent configurations.
\end{enumerate}
Note that unlike the deterministic dynamics~(\ref{eq:eom-precessional}),
the above Monte Carlo scheme does not conserve energy.

\subsection{Numerical results}
\label{sec:numerics-1}

\begin{figure}
\centering
        \includegraphics[width=7cm]{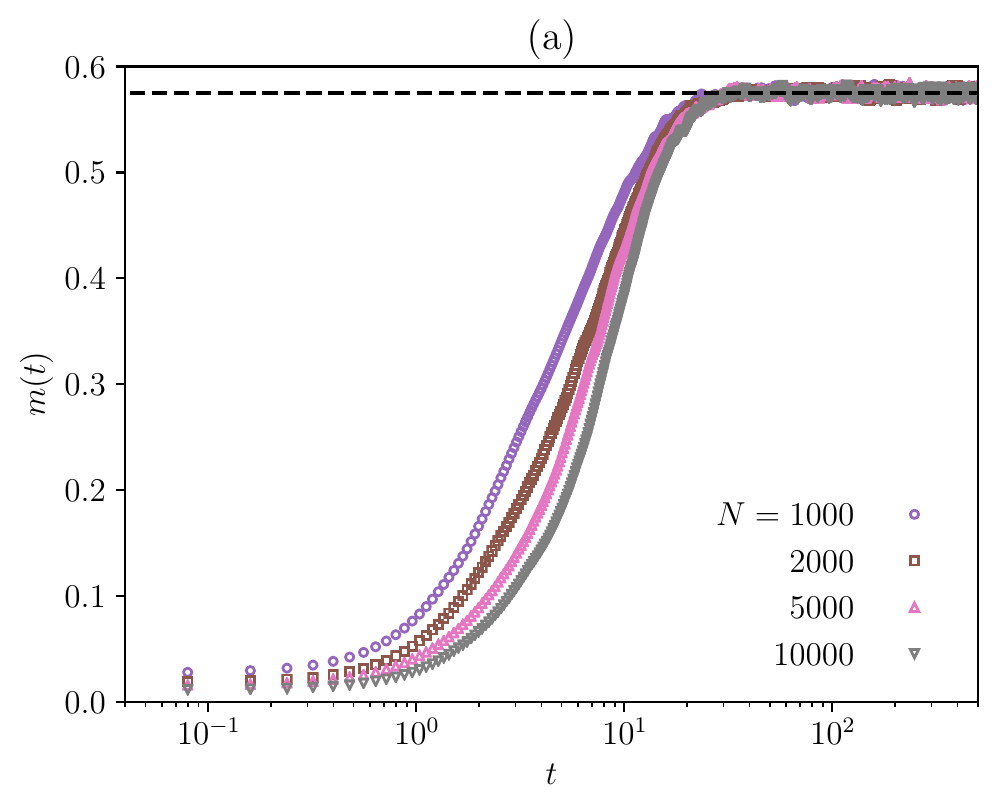}
        \includegraphics[width=7cm]{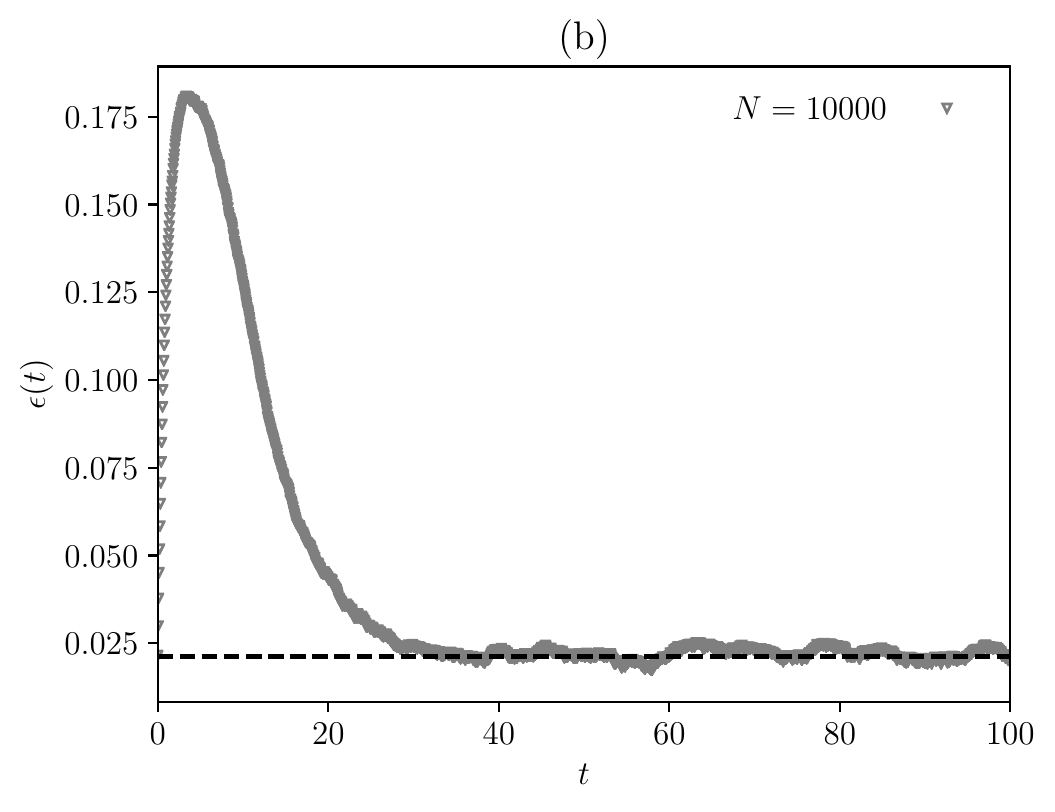}
        \includegraphics[width=7cm]{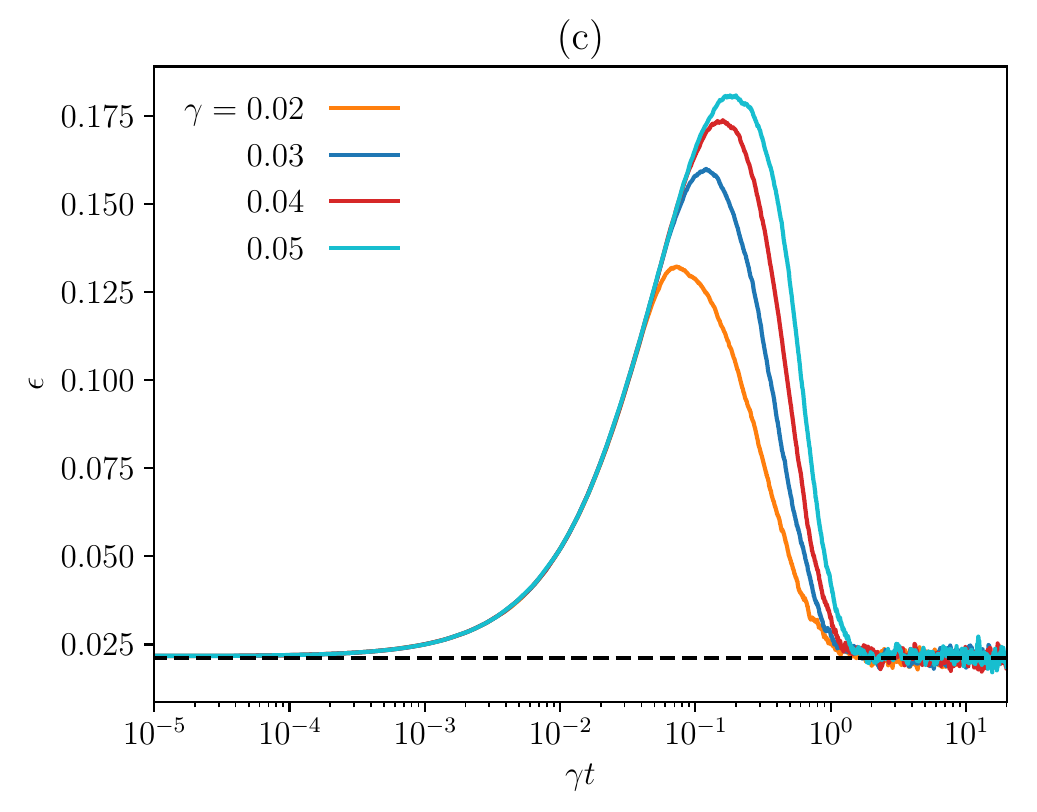}
\caption{For the same parameter values as in
Fig.~\ref{fig:s2-fig1-nonoise}, the points in
panel (a) are obtained from numerical integration of the stochastic
dynamics~(\ref{eq:eom-noise}) for $\gamma=0.05$ and at a value of
temperature that ensures that one obtains the same value of the
equilibrium magnetization as that obtained at the value of energy chosen
for the deterministic dynamics studied in Fig.~\ref{fig:s2-fig1-nonoise}. The results suggest a
fast relaxation to equilibrium on a size-independent timescale, with no
sign of quasistationarity. In panel (b), we show for the parameter
values of panel (a) that indeed at the studied value of temperature, the
average energy of the system in equilibrium matches up to numerical
accuracy the conserved value
of energy (dashed line) chosen for the deterministic dynamics in
Fig.~\ref{fig:s2-fig1-nonoise}. The
figure in panel (c) shows for $N=10000$ the evolution of energy under the
stochastic dynamics~(\ref{eq:eom-noise}) for four values of the
dissipation parameter $\gamma$. Scaling collapse of the data suggests
relaxation of the initial state over the timescale $\sim 1/\gamma$. Here
the dashed line corresponds to the initial and the final energy value.
Note that it is the final energy value that gets fixed in our numerical
scheme by our choice of the temperature, while the choice of the initial
energy is immaterial as it anyways is not conserved by the dynamics and
will evolve to the allowed final value. In the plot of panel (c), the
initial energy value happens to have been chosen to equal the final
allowed value.}
\label{fig:s2-fig2}
\end{figure}

\begin{figure}
\centering
        \includegraphics[width=7cm]{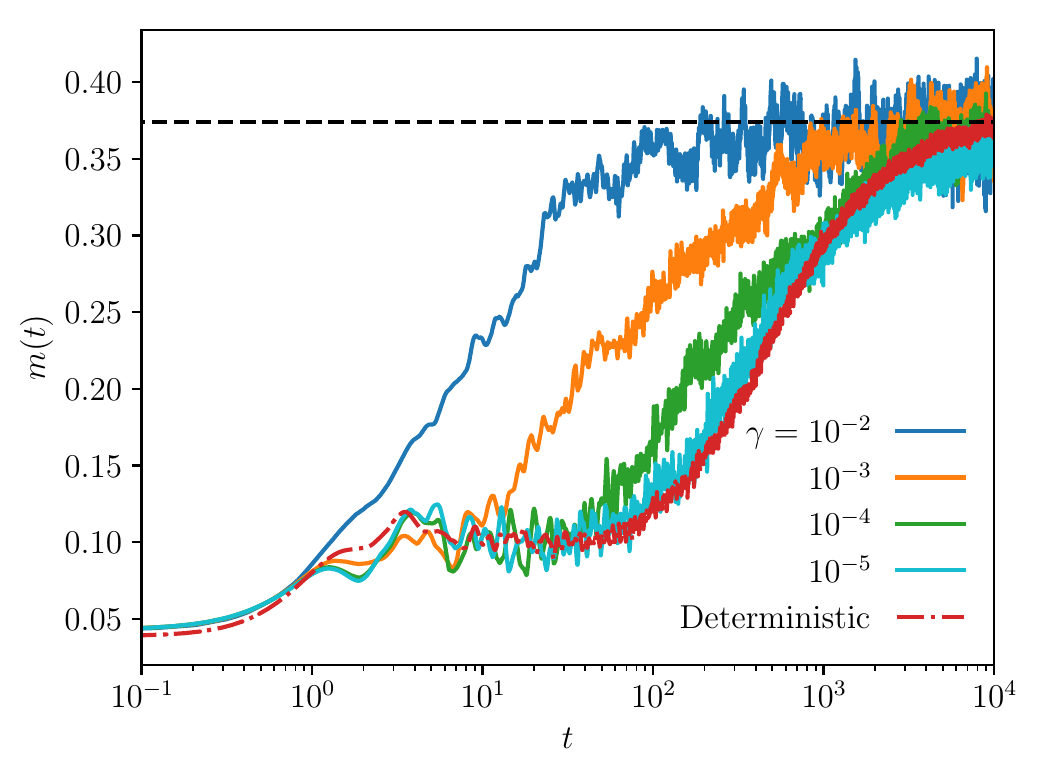}
\caption{Considering the stochastic
dynamics~(\ref{eq:eom-noise}) with $n=1,D=5.0,N=500$, four values
of $\gamma$, and with (\ref{eq:FD}) as the initial state with $\beta_{\rm FD}=1000$, initial energy $\epsilon=0.1473$, the
figure shows a cross-over in the relaxation behavior, from
a fast to a slow one, as one tunes the parameter $\gamma$ from high to
low values. Here, we keep the temperature fixed to a value such that one obtains the same value of the equilibrium magnetization as
that obtained for the deterministic dynamics~(\ref{eq:eom-precessional})
with energy equal to $\epsilon$ whose results are also
included in the plot. The observed behavior is consistent with the conclusion drawn in
Section~\ref{sec:noisy-finiteN}. Here, the dashed
line represents the value of equilibrium magnetization at the studied
temperature. }
\label{fig:s2-cross-over}
\end{figure}

Here, we first discuss for $n=1$ results
obtained from numerical integration of the stochastic
dynamics~(\ref{eq:eom-noise}) on implementing the algorithm discussed in
Appendix C. For the results reported in
this work, we take $\gamma=0.05$ and integration timestep equal to $10^{-3}$. Data averaging has been typically over several hundreds to
thousand runs of the dynamics starting from different realizations of
the FD state~(\ref{eq:FD}). 
Our aim here is to compare stochastic dynamics
results with those from deterministic dynamics observed at a given
energy $\epsilon$. By virtue of equivalence of
microcanonical and canonical ensembles in equilibrium, we choose the
temperature $T$ in the stochastic dynamics to have a value that ensures
that one has in equilibrium the same value of magnetization as the one
observed for the deterministic dynamics at energy $\epsilon$; this is
done by using plots such as those in Fig.~\ref{fig:eTm}.
Figure~\ref{fig:s2-fig2}(a) shows that under stochastic dynamics with
$1/\gamma \ll N$, the
initial FD state shows a fast relaxation to Boltzmann-Gibbs equilibrium on a
size-independent timescale and there is no sign of quasistationarity
during the process of relaxation. Figure~\ref{fig:s2-fig2}(b) shows that
at the chosen value of $T$, the average energy of the system in
equilibrium does coincide with the conserved energy of the deterministic
dynamics, as it should due to our choice of $T$.
Figure~\ref{fig:s2-fig2}(c) shows for $N=10000$ the evolution of energy under the
stochastic dynamics~(\ref{eq:eom-noise}) for four values of the
dissipation parameter $\gamma$. Scaling collapse of the data suggests
relaxation of the initial state over the timescale $\sim 1/\gamma$,
consistent with our analysis in Section~\ref{sec:noisy-finiteN}. Relaxation on a
size-independent timescale is also observed for energies $\epsilon^\star
< \epsilon < \epsilon_c$ (see Fig.~\ref{fig:s2-fig3-nonoise}(a)) and for
energies $\epsilon > \epsilon_c$ (see
Fig.~\ref{fig:s2-fig4-nonoise}(a)); note that in these cases too we have
$1/\gamma \ll N$. The expected cross-over in the relaxation behavior as
one tunes for a fixed $N$ the value of $\gamma$ from low to high values
is verified by the plot in Fig.~\ref{fig:s2-cross-over}. Similar results
as for $n=1$ are also observed for $n=2$, see
Figs.~\ref{fig:s4-fig2},~\ref{fig:s4-fig3-nonoise},
and~\ref{fig:s4-fig4-nonoise}.

\begin{figure}[!ht]
\centering
        \includegraphics[width=7cm]{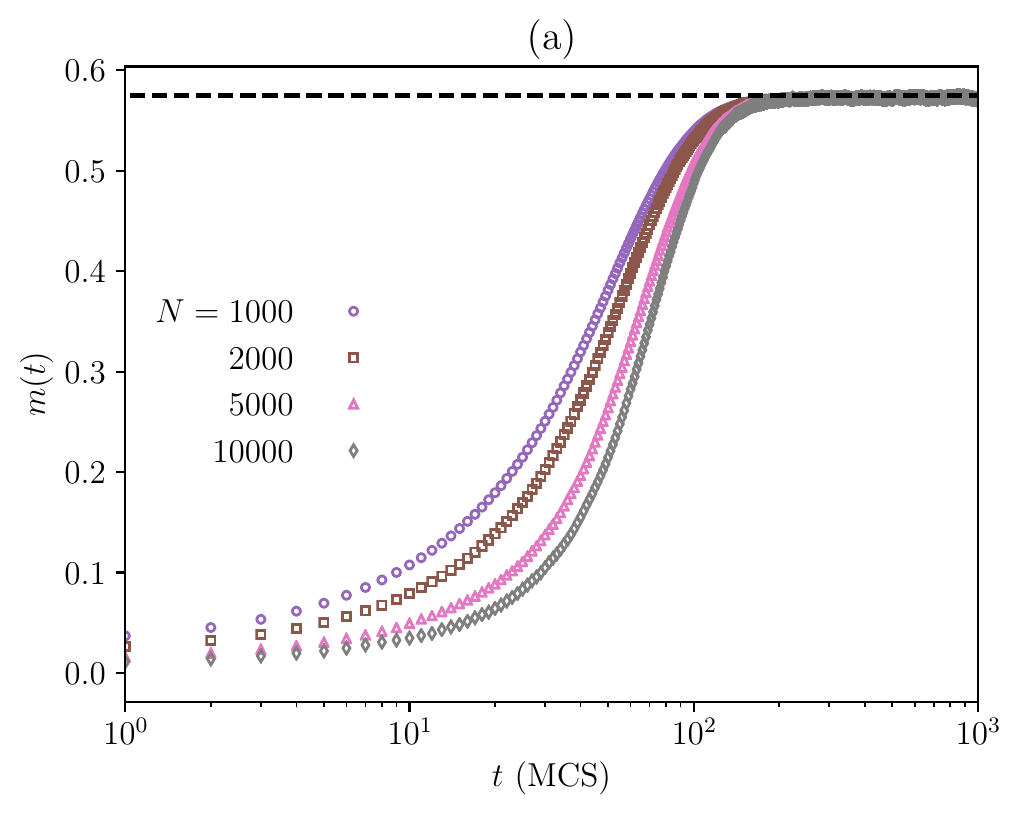}
        \includegraphics[width=7cm]{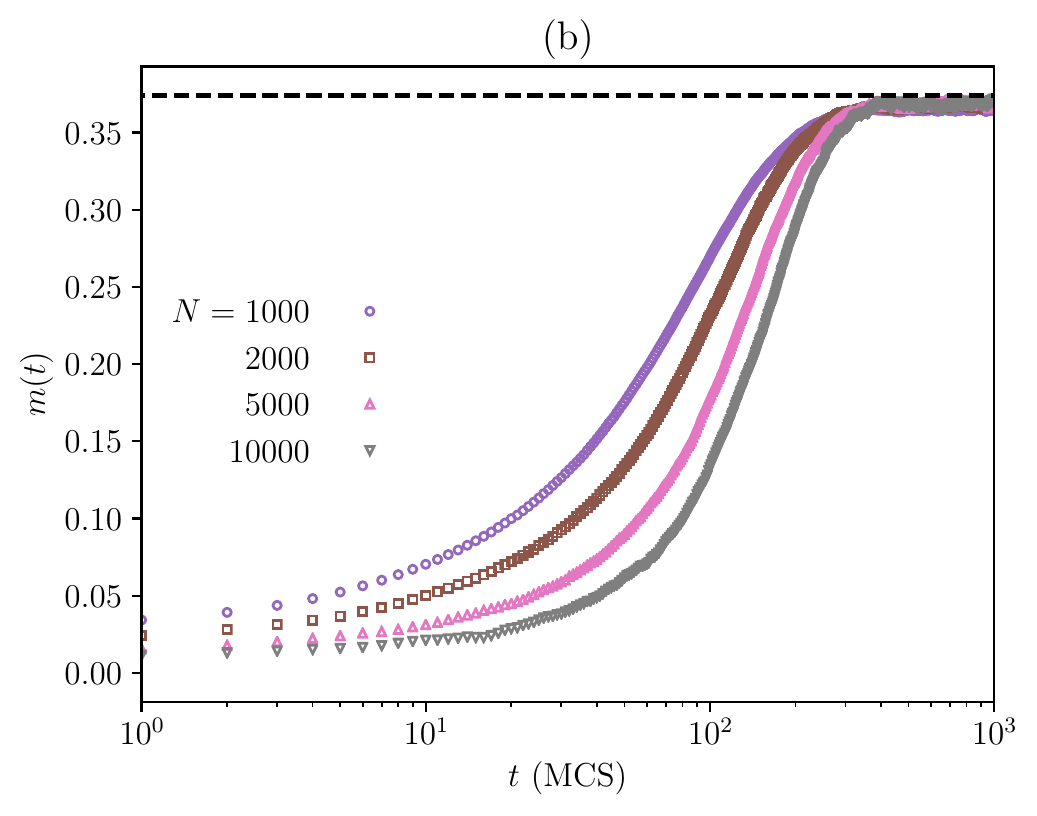}
        \caption{For the model~(\ref{eq:H}) with $n=1$ and $D=5.0$, the figure shows the relaxation under Glauber
Monte Carlo dynamics of an initial
nonmagnetized FD
state~(\ref{eq:FD}) with $\beta_{\rm FD}=1000$ for (a) the same temperature as in
Fig.~\ref{fig:s2-fig2}(c) and (b) the same temperature as in
Fig.~\ref{fig:s2-fig3-nonoise}(a). Here the dashed line denotes the equilibrium
magnetization value at the temperature at which the Monte Carlo dynamics
is implemented. The figures suggest the absence of any
quasistationary behavior and a fast relaxation on a size-independent
timescale to equilibrium. We have observed similar fast relaxation also
for temperatures corresponding to energies $\epsilon>\epsilon_c$ of the
deterministic dynamics (data not shown here).}
\label{fig:s2-fig5}
\end{figure}

\begin{figure}[!ht]
\centering
        \includegraphics[width=7cm]{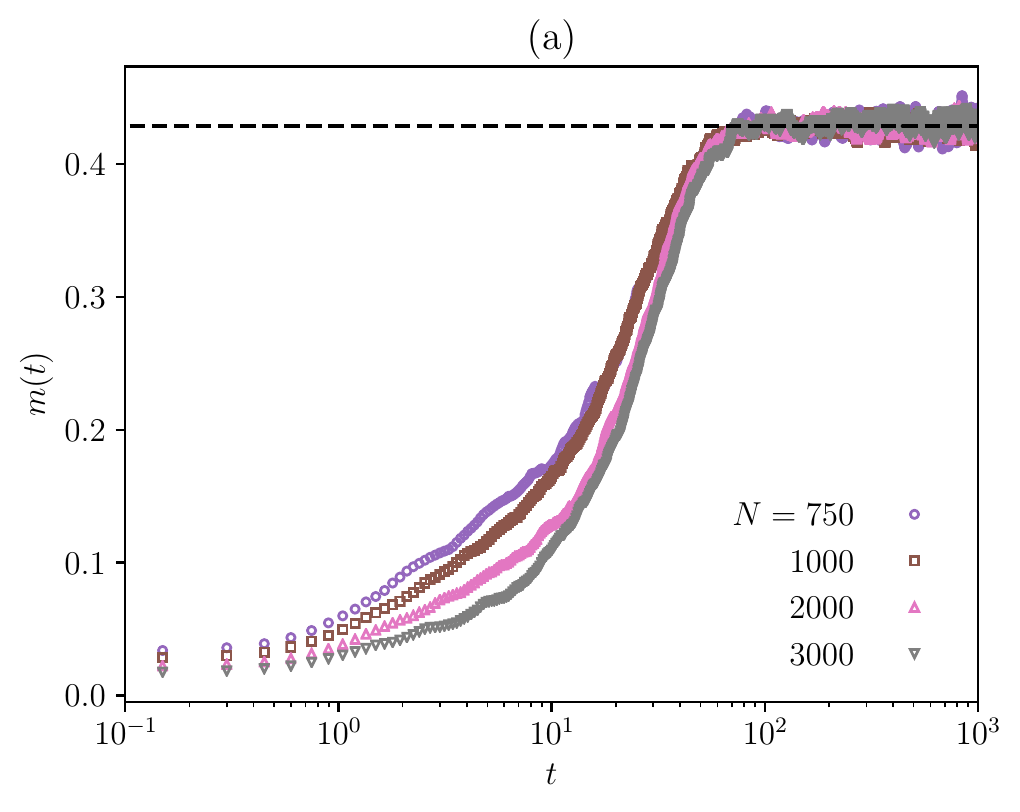}
        \includegraphics[width=7cm]{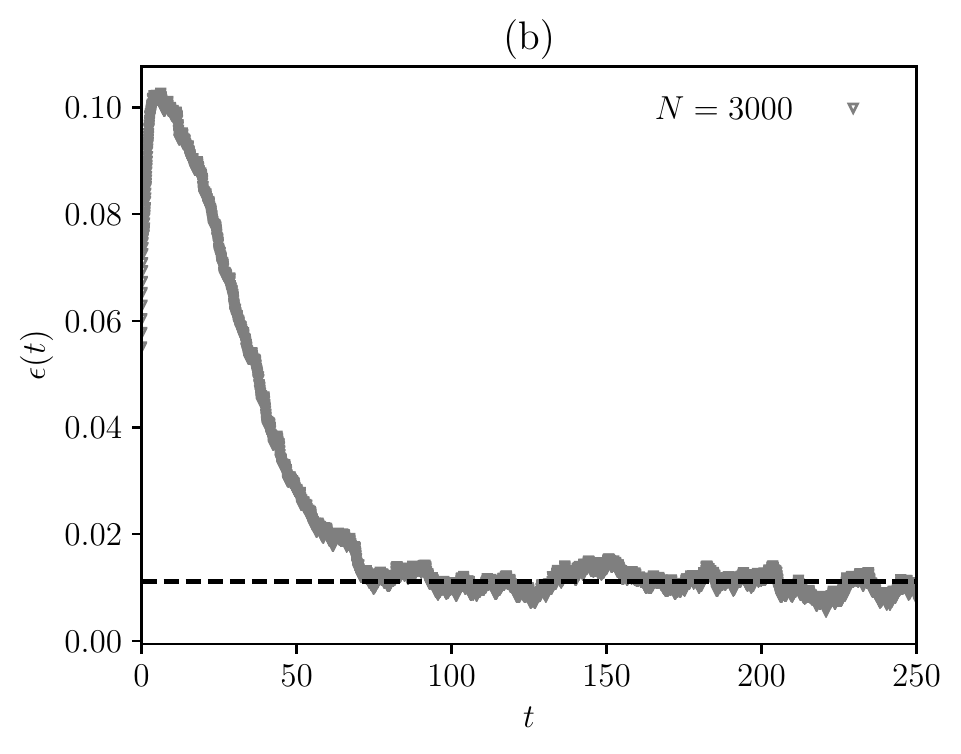}
        \includegraphics[width=7cm]{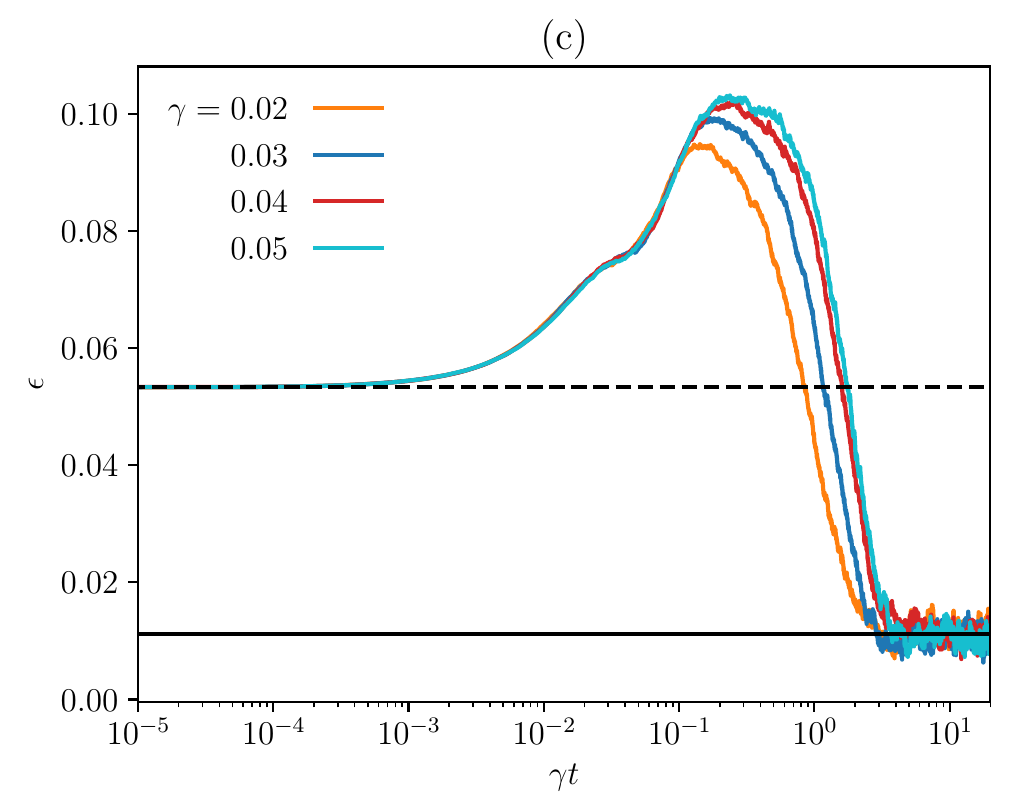}
\caption{For the same parameter values as in
Fig.~\ref{fig:s4-fig1-nonoise}, the points in
panel (a) are obtained from numerical integration of the stochastic
dynamics~(\ref{eq:eom-noise}) for $\gamma=0.05$ and at a value of
temperature that ensures that one obtains the same value of the
equilibrium magnetization as that obtained at the value of energy chosen
for the deterministic dynamics studied in
Fig.~\ref{fig:s4-fig1-nonoise}. The results suggest a
fast relaxation to equilibrium on a size-independent timescale, with no
sign of quasistationarity. In panel (b), we show for the parameter
values of panel (a) that indeed at the studied value of temperature, the
average energy of the system in equilibrium matches the conserved value
of energy (dashed line) chosen for the deterministic dynamics in Fig.~\ref{fig:s4-fig1-nonoise}.
The figure in panel (c) shows for $N=3000$ the evolution of energy under the
stochastic dynamics~(\ref{eq:eom-noise}) for four values of the
dissipation parameter $\gamma$. Scaling collapse of the data suggests
relaxation of the initial state over the timescale $\sim 1/\gamma$.
Here, the dashed line (respectively, the solid line) corresponds to
initial (respectively, final) energy value.}
\label{fig:s4-fig2}
\end{figure}

\begin{figure}
\centering
        \includegraphics[width=7cm]{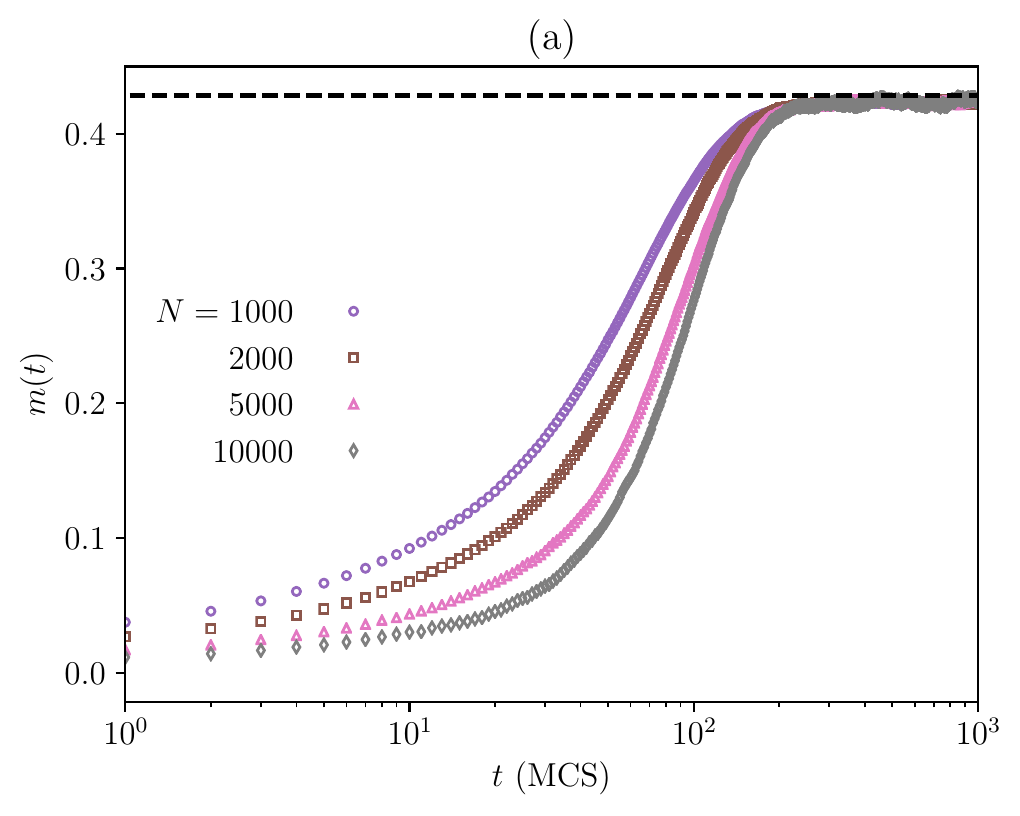}
        \includegraphics[width=7cm]{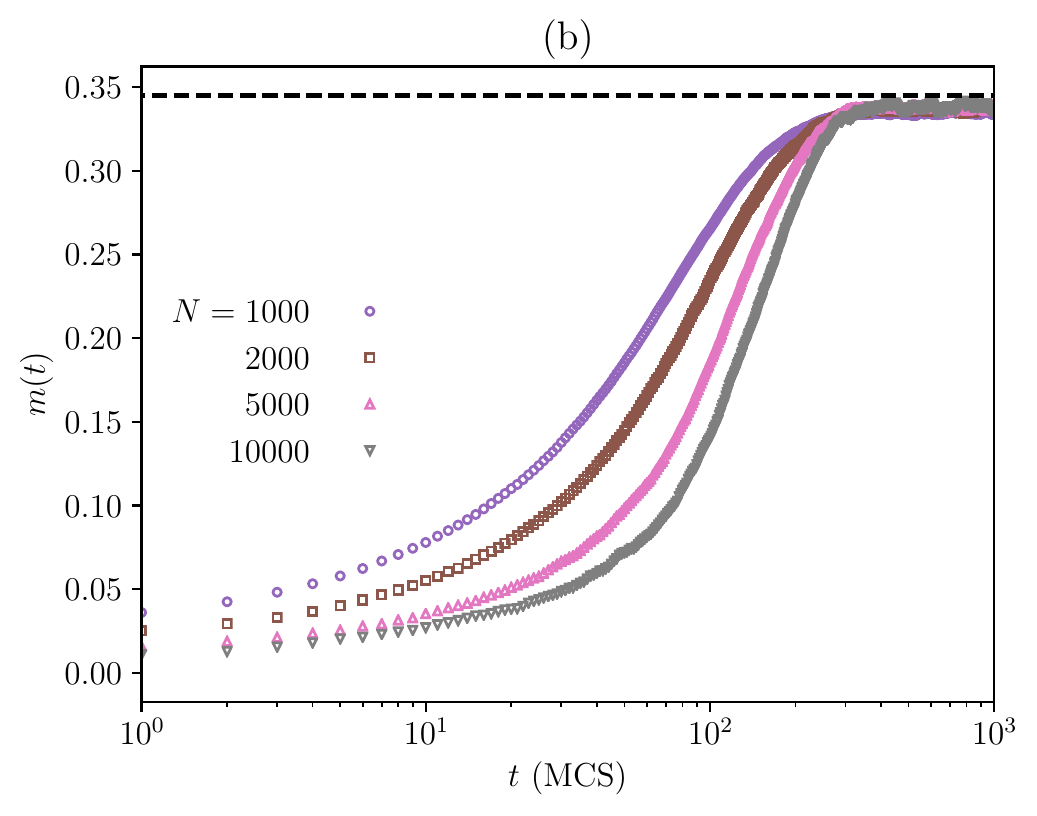}
\caption{For the model~(\ref{eq:H}) with $n=2$ and $D=15.0$ thus yielding
$\epsilon_c\approx 0.1175$, the figure shows the relaxation under Glauber
Monte Carlo dynamics of an initial
nonmagnetized FD
state~(\ref{eq:FD}) with $\beta_{\rm FD}=100$ for (a) the same temperature as in
Fig.~\ref{fig:s4-fig2}(c) and (b) the same temperature as in
Fig.~\ref{fig:s4-fig3-nonoise}(a). Here the dashed line denotes the equilibrium
magnetization value at the temperature at which the Monte Carlo dynamics
is implemented. The figures suggest the absence of any
quasistationary behavior and a fast relaxation on a size-independent
timescale to equilibrium. We have observed similar fast relaxation also
for temperatures corresponding to energies $\epsilon>\epsilon_c$ of the
deterministic dynamics (data not shown here).}
\label{fig:s4-fig5}
\end{figure}

Next, we show in Figs.~\ref{fig:s2-fig5} and~\ref{fig:s4-fig5} the results from Glauber
Monte Carlo simulation of the system~(\ref{eq:H}) for $n=1,2$. Here too
one observes a fast relaxation to equilibrium over a size-independent
time scale. This may be explained based on the fact that size-dependent
relaxation is a feature of energy-conserving Vlasov dynamics, as is
evident from the discussions in Section~\ref{sec:noiseless}. While within the
scheme~(\ref{eq:eom-noise}), energy conservation is violated on the
timescale $\sim 1/\gamma$, the same within the Monte Carlo scheme
happens on the scale of one time unit. Hence, obviously, within the
later scheme, size-dependent relaxation will not be observed anyhow,
while in the former, fast
relaxation requires choosing the noise to be strong enough that
$1/\gamma \ll N$.

\section{Conclusions}
\label{sec:conclusions}

In this work, we wanted to assess the effects of stochasticity, such as
those arising from the finiteness of system size or those due to
interaction with the external environment, on the relaxation properties
of a model long-range interacting system of classical Heisenberg spins. Under deterministic spin precessional dynamics,
we showed for a wide range of energy values a slow relaxation to
Boltzmann-Gibbs equilibrium over a timescale that diverges with the
system size. The corresponding stochastic dynamics, modeling interaction
with the environment and constructed in the spirit of (i) the stochastic
Landau-Lifshitz-Gilbert equation, and (ii) the Glauber Monte Carlo dynamics, however shows a fast relaxation to equilibrium on a size-independent
timescale, with no signature of quasistationarity. Our work establishes
unequivocally how quasistationarity observed in deterministic dynamics
of long-range systems is washed away by fluctuations
induced through contact with the environment.

In the light of results on slow relaxation to equilibrium reported in
this work, it would be interesting to address the issue of how the
system~(\ref{eq:H}) prepared either in Boltzmann-Gibbs equilibrium or in QSSs responds to an external
field. One issue of particular relevance is when the field is small, and
one has for short-range systems in equilibrium a linear
response to the field that may be expressed in terms of fluctuation properties of the system
in equilibrium. While investigation of similar fluctuation-response
relations for LRI systems has been pursued in the context of particle
dynamics (e.g., that of the HMF model~\cite{Patelli:2012,Ogawa:2012})
and strange scaling of fluctuations in finite-size systems has
been reported~\cite{Yamaguchi:2016}, it would be interesting to pursue such a study for the spin
model~(\ref{eq:H}). Investigations in this direction have been reported
in Ref.~\cite{Yamaguchi:2019}.

\section{Acknowledgements}
The work of DD is supported by UGC-NET Research Fellowship Sr.
No. 2121450744, dated 29-05-2015, Ref. No. 21/12/2014(ii) EU-V. SG thanks
Yoshiyuki Y. Yamaguchi for fruitful discussions, for suggesting to study the
model~(\ref{eq:H}) for $n=2$, and for pointing out
Ref.~\cite{Kventsel:1984}. The
manuscript was finalized while S.G. was visiting the International Centre for Theoretical Physics (ICTP), Trieste during May 2019 as a Regular Associate of the Quantitative Life Sciences section, and would like to acknowledge the support and hospitality of the ICTP.  

\section{Appendix A: Derivation of Eqs.~(\ref{eq:FD-norma}),~(\ref{energy}),
and~(\ref{eq:spin4-FD-stability-criterion}) of the main text}
\label{appendix1}

The normalization $A$ satisfies
\begin{eqnarray}
 1&=& A \int^{\pi}_{0} {\rm d\theta} \sin\theta \frac{1}{1+e^{\beta_{\rm
 FD}(\cos^2\theta - \mu)}}\nonumber \\
&=&2A \Big[ \frac{1}{1+e^{\beta_{\rm FD}(1-\mu)}}  + \int^{1}_{0}   \sqrt{x}
\Big( - \frac{\partial}{\partial x} f_{\rm FD}\Big)\Big],
\label{eq:FD-norm-calculation}
\end{eqnarray}
with $f_{\rm FD}(x)=1/(1+e^{\beta_{\rm FD}(x-\mu)})$, and where in obtaining the last equality, we have performed integration by
parts.
When $\beta_{\rm FD}$ is large, the first term on the right hand side of
Eq.~(\ref{eq:FD-norm-calculation}) drops out. In order to evaluate the
second term, using the fact for large $\beta_{\rm FD}$, $\partial f_{\rm
FD}(x)/\partial x =-\delta(x-\mu)$,  we Taylor expand $\sqrt{x}$ about $\mu$,
which on substituting in Eq.~(\ref{eq:FD-norm-calculation}) gives for large
$\beta_{\rm FD}$ that
\begin{eqnarray}
2A \Big[  \sqrt{\mu} \bar{I}_0 +  \frac{1}{2\beta_{\rm FD}\sqrt{\mu}} \bar{I}_1 -
\frac{1}{8\beta_{\rm FD}^2 \mu^{3/2}} \bar{I}_2  \Big] = 1,
\label{eq:FD-norm-calculation-1}
\end{eqnarray}
with
\begin{eqnarray}
\bar{I}_0&=& \int^{1}_0 {\rm d}x \Big( - \frac{\partial}{\partial x}
f_{\rm FD}(x) \Big)
\approx \int^{\infty}_{-\infty} {\rm d}y \frac{e^y}{(1+e^y)^2} = 1,
\nonumber \\ 
\bar{I}_1&=& \int^{1}_0 {\rm d}x ~\beta_{\rm FD}(x-\mu) \Big( -
\frac{\partial}{\partial x} f_{\rm FD}(x) \Big)
\approx \int^{\infty}_{-\infty} {\rm d}y \frac{ye^y}{(1+e^y)^2} = 0,
\nonumber \\
\bar{I}_2&=& \int^{1}_0 {\rm d}x ~\beta_{\rm FD}^2(x-\mu)^2 \Big( -
\frac{\partial}{\partial x} f_{\rm FD}(x) \Big)\approx \int^{\infty}_{-\infty} {\rm d}y \frac{y^2e^y}{(1+e^y)^2} =
\frac{\pi^2}{3}. \nonumber \\
\label{eq:FD-integrals}
\end{eqnarray}
Here, we have considered the limit of large $\beta_{\rm FD}$ in evaluating all
the three integrals $\bar{I}_0,\bar{I}_1,\bar{I}_2$.
Using Eq.~(\ref{eq:FD-integrals}) in
Eq.~(\ref{eq:FD-norm-calculation-1}), we obtain for large $\beta_{\rm
FD}$ the
following result correct to order $1/\beta_{\rm FD}^2$:
\begin{equation}
A = \frac{1}{2\sqrt{\mu}} \Big[ 1+ \frac{\pi^2}{24\beta_{\rm FD}^2 \mu^2} \Big], 
\label{eq:ap-FD-norma}
\end{equation}
which is Eq.~(\ref{eq:FD-norma}) of the main text.

The energy corresponding to the state~(\ref{eq:FD}) is given by 
\begin{eqnarray}
\epsilon &=& \frac{AD}{2\pi} \int^{\pi}_{0} \sin\theta {\rm d}\theta
{\rm d}\phi  ~\frac{\cos^{2n} \theta}{1+e^{\beta_{\rm FD}(\cos^2\theta - \mu)}}\nonumber \\
&=& \frac{2AD}{2n+1} ~\Big[ \frac{1}{1+e^{\beta_{\rm FD}(1 - \mu)}}+ \int^{1}_0
{\rm d}x ~x^{(2n+1)/2} \Big( - \frac{\partial}{\partial x} f_{\rm
FD}(x) \Big)\Big], 
\label{eq:FD-energy-1}
\end{eqnarray}
where we have used
integration by parts to arrive at the last equality. 
For large $\beta_{\rm FD}$, the first term on the right hand side of
Eq.~(\ref{eq:FD-energy-1}) drops out, while noting that we have $\partial f_{\rm FD}(x)/\partial x =-\delta(x-\mu)$, we evaluate
the second term by Taylor expanding $x^{(2n+1)/2}$ about $x=\mu$. We
finally get for
large $\beta_{\rm FD}$ that
\begin{eqnarray}
&&\fl \epsilon = \frac{2AD}{2n+1} \Big[ \mu^{(2n+1)/2} \bar{I}_0 +
\frac{(2n+1)\mu^{(2n-1)/2}}{2\beta_{\rm FD}} \bar{I}_1+ \frac{(2n+1)(2n-1)\mu^{(2n-3)/2}}{8\beta_{\rm FD}^2} \bar{I}_2 +\cdots \Big],
\end{eqnarray}
with $\bar{I}_0,\bar{I}_1,\bar{I}_2$ given by
Eq.~(\ref{eq:FD-integrals}). Using the latter, we get for large
$\beta_{\rm FD}$
that
\begin{equation}
\epsilon = \frac{2AD}{2n+1} \Big[ \mu^{(2n+1)/2} +
\frac{(2n+1)(2n-1)\pi^2 \mu^{(2n-3)/2}}{24\beta_{\rm FD}^2} \Big],
\end{equation}
correct to order
$1/\beta_{\rm FD}^{2}$. On using Eq~(\ref{eq:ap-FD-norma}), we finally get 
\begin{equation}
\epsilon = \frac{D}{2n+1} \Big[\mu^{2n/2} +
\frac{(2n)^2\pi^2}{24\beta_{\rm FD}^2} \mu^{(2n-4)/2} \Big], 
\label{app-energy}
\end{equation}
correct to order $1/\beta_{\rm FD}^2$.
Equation~(\ref{app-energy}) is Eq.~(\ref{energy}) of the main text.

Our next job is to show how using Eq.~(\ref{eq:FD}) in
Eq.~(\ref{eq:stability-condition}) leads to 
Eq.~(\ref{eq:spin4-FD-stability-criterion}) of the main text.
It may be straightforwardly shown by using Eqs.~(\ref{eq:FD}) and 
Eq.~(\ref{eq:stability-condition}) that  
\begin{equation}
1=2nDA \int^{1}_{0} {\rm d}x
~g(x)\Big[ -
\frac{\partial}{\partial x} f_{\rm FD} \Big];~~
g(x)=\frac{x^{(2n-1)/2}-x^{(2n+1)/2}}{(2n)^2D^2x^{2n-1}-\omega^2}.
\label{eq:FD-w-1}
\end{equation}
Noting that for large $\beta_{\rm FD}$, one has $\partial f_{\rm FD}(x)/\partial
x=-\delta(x-\mu)$, we may expand $g(x)$ in a Taylor series about
$x=\mu$, and evaluate the right hand side. One gets to order
$1/\beta_{\rm FD}^2$ the result
\begin{equation}
 2nDA \Big[g(\mu)+  \frac{g''(\mu)}{2\beta_{\rm FD}^2} \frac{\pi^2}{3} \Big]  = 1.
\end{equation}
Using Eq.~(\ref{eq:ap-FD-norma}), we get to order $1/\beta_{\rm FD}^2$ the equation
\begin{equation}
g(\mu)\mu^{-1/2}+  \frac{\pi^2}{24\beta_{\rm FD}^2} \Big[ g(\mu) \mu^{-5/2} + 4
g''(\mu) \mu^{-1/2} \Big] = \frac{1}{nD},
\label{eq:app-spin4-FD-stability-criterion}
\end{equation}
where $\mu$ is to be obtained by solving Eq.~(\ref{app-energy}).
Equation~(\ref{eq:app-spin4-FD-stability-criterion}) is
Eq.~(\ref{eq:spin4-FD-stability-criterion}) of the main text.

\section{Appendix B: Derivation of Eqs.~(\ref{eq:Vlasov-1}) and
(\ref{eq:time-evolution-stochastic}) of the main
text}
\label{appendix2}

Here, we derive Eqs.~(\ref{eq:Vlasov-1}) and
(\ref{eq:time-evolution-stochastic}) of the main text. We start with the
equation of motion~(\ref{eq:eom-noise}): 
\begin{equation}
\dot{\bf S}_{i}={\bf S}_{i}\times({\bf
h}_{i}^{{\rm eff}}+\boldsymbol{\eta}_{i}(t))-\gamma{\bf
S}_{i}\times({\bf S}_{i}\times({\bf h}_{i}^{{\rm
eff}}+\boldsymbol{\eta}_{i}(t))),\label{eq:eom-sllg}
\end{equation}
where $\boldsymbol{\eta}_{i}(t)$ is a Gaussian white noise with 
\begin{equation}
\langle\eta_{i\alpha}(t)\rangle=0;\thinspace\thinspace\langle\eta_{i\alpha}(t)\eta_{j\beta}(t')\rangle=2{\cal
D}\delta_{ij}\delta_{\alpha\beta}\delta(t-t'),\label{eq:noise}
\end{equation}
and
\begin{equation}
{\bf h}_i^{\rm
eff}={\bf m}+{\bf h}_i^{\rm aniso};~{\bf h}_i^{\rm
aniso}=(0,0,-2nDS_{iz}^{2n-1}).
\end{equation}
In terms of components, Eq.\,(\ref{eq:eom-sllg}) reads 
\begin{equation}
\dot{S_{i}^{\alpha}}=f_{i}^{\alpha}({{\bf S}_{i}})+g_{i}^{\alpha\lambda}(\{{\bf S}_{i}\})\eta_{i}^{\lambda},
\label{eq:fi-gi-defn}
\end{equation}
where we have used Einstein summation convention for repeated indices,
and
\begin{eqnarray}
&&f_{i}^{\alpha}(\{{\bf S}_{i}\})=
\epsilon_{\alpha\beta\lambda}S_{i}^{\beta}h_{i}^{{\rm eff},\lambda}
-\gamma \epsilon_{\alpha\beta\lambda} \epsilon_{\lambda\sigma\rho}
S_{i}^{\beta} S_{i}^{\sigma} h_{i}^{{\rm eff},\rho}, \\
&& g_{i}^{\alpha\beta}(\{{\bf S}_{i}\})=
\epsilon_{\alpha\lambda\beta}S_{i}^{\lambda}-\gamma
\epsilon_{\alpha\lambda\rho} \epsilon_{\rho\sigma\beta} S_{i}^{\lambda}
S_{i}^{\sigma}. \label{eq:g}
\end{eqnarray}

Let us define $F_d({\bf S},t)$, the discrete single-spin density
function, as
\begin{equation}
F_d({\bf S},t)=\frac{1}{N}\sum_{i=1}^N \delta({\bf S}-{\bf S}_i(t)).
\end{equation}
For a given noise
realization $\{\bfeta_i\}$, let us obtain the time evolution equation
for $F_d$. To this end, differentiating both sides of the last equation with respect to time, using
Eq.~(\ref{eq:fi-gi-defn}), and the property $a\delta(a-b)=b\delta(a-b)$,
one gets
\begin{equation}
\frac{\partial}{\partial t}F_d({\bf S},t)=-\frac{\partial}{\partial
S^{\alpha}}\Big[\Big(f^{\alpha}+g^{\alpha\lambda}\eta^{\lambda}\Big)F_d({\bf
S},t)\Big],
\label{eq:fd-eqn}
\end{equation}
with
\begin{eqnarray}
&&\fl f^{\alpha}({\bf S})=
\epsilon_{\alpha\beta\lambda}S^{\beta}h^{{\rm eff},\lambda}
-\gamma \epsilon_{\alpha\beta\lambda} \epsilon_{\lambda\sigma\rho}
S^{\beta} S^{\sigma} h^{{\rm eff},\rho}; \\
&&\fl {\bf h}^{\rm eff}\equiv {\bf h}^{\rm eff}[F_d]={\bf
m}[F_d]+(0,0,-2nDS_z^{2n-1});~~{\bf m}[F_d]\equiv\int {\rm d}{\bf
S}~{\bf S}F_d({\bf
S},t), \\
&& \fl g^{\alpha\beta}({\bf S})=
\epsilon_{\alpha\lambda\beta}S^{\lambda}-\gamma
\epsilon_{\alpha\lambda\rho} \epsilon_{\rho\sigma\beta} S^{\lambda}
S^{\sigma}. \label{eq:g1}
\end{eqnarray}

Averaging Eq.~(\ref{eq:fd-eqn}) over the noise statistics (\ref{eq:noise}), one gets for
the averaged distribution $P_d({\bf S},t)$
the equation~\cite{Kubo:1963,Nishino:2015} 
\begin{equation}
\frac{\partial P_d({\bf S},t)}{\partial t}  =-\frac{\partial}{\partial
S^{\alpha}}\Big[f^{\alpha}P_d-{\cal
D}g^{\alpha\beta}\frac{\partial}{\partial
S^{\lambda}}(g^{\lambda\beta}P_d)\Big].\label{eq:fpe}
\end{equation}
Using Eq.~(\ref{eq:g1}), we have 
\begin{equation}
\frac{\partial}{\partial S^{\alpha}}g^{\alpha\lambda}  =\epsilon_{\alpha\beta\lambda}\delta_{\alpha\beta}-\gamma\delta_{\alpha\alpha}S^{\lambda}-\gamma S^{\alpha}\delta_{\alpha\lambda}=-4\gamma S^{\lambda},
\end{equation}
so that 
\begin{equation}
g^{\alpha\beta}\frac{\partial}{\partial S^{\lambda}}g^{\lambda\beta}  =-4\gamma\Big(\epsilon_{\alpha\lambda\beta}S^{\lambda}-\gamma S^{\alpha}S^{\beta}+\gamma\delta_{\alpha\beta}\Big)S^{\beta}=0,
\end{equation}
where we have used the fact that $\epsilon_{\alpha\lambda\beta}$
is completely antisymmetric with respect to the indices. Consequently,
Eq.~(\ref{eq:fpe}) gives 
\begin{equation}
\frac{\partial P_d({\bf S},t)}{\partial t} 
=-\frac{\partial}{\partial
S^{\alpha}}\Big[f^{\alpha}-{\cal
D}g^{\alpha\beta}g^{\lambda\beta}\frac{\partial}{\partial
S^{\lambda}}\Big]P_d.
\label{eq:fpe1}
\end{equation}
Next, Eq.~(\ref{eq:g1}) gives
$g^{\alpha\beta}g^{\lambda\beta}=(1+\gamma^2)
\epsilon_{\alpha\sigma\beta} \epsilon_{\lambda\rho\beta} S^{\sigma}
S^{\rho}$, so that the right hand side of Eq.~(\ref{eq:fpe1}) now reads
\begin{eqnarray}
&&\fl -\frac{\partial}{\partial
S^{\alpha}} \Big[ \Big( \epsilon_{\alpha\beta\lambda}S^{\beta}h^{{\rm
eff},\lambda} -\gamma  \epsilon_{\alpha\beta\lambda}
\epsilon_{\lambda\sigma\rho} S^{\beta} S^{\sigma} h^{{\rm eff},\rho}  -
{\cal D}(1+\gamma^{2})  \epsilon_{\alpha\sigma\beta}
\epsilon_{\lambda\rho\beta} S^{\sigma} S^{\rho}
\frac{\partial}{\partial S^{\lambda}}  \Big)  P_d\Big].
\end{eqnarray}
Consequently, Eq.~(\ref{eq:fpe1}) now reads
\begin{eqnarray}
&&\fl 
\frac{\partial P_d({\bf S},t)}{\partial t}= -\frac{\partial}{\partial{\bf
 S}}\cdot\Big[({\bf S}\times{\bf h}^{{\rm
 eff}})P_d-\gamma(({\bf
 S}\times{\bf S}\times{\bf h}^{{\rm eff}})P_d)+{\cal
 D}(1+\gamma^{2})({\bf
 S}\times({\bf S}\times\frac{\partial}{\partial{\bf
 S}}))P_d\Big],\nonumber \\\label{eq:FPE1}
\end{eqnarray}
where note that $\partial/\partial{\bf
 S}\cdot({\bf S}\times{\bf h}^{{\rm
 eff}})P_d=({\bf S}\times{\bf h}^{{\rm
 eff}})\cdot \partial P_d/\partial {\bf S}$.

Let us define an averaged one-spin density function $P({\bf S},t)$ as
the average of $P_d({\bf S},t)$ with respect to a large number of
initial conditions close to the same macroscopic state. To this end, we
have to leading order the expansion
\begin{equation}
P_d({\bf S},t)=P_0({\bf S},t)+\frac{1}{\sqrt{N}}\delta P({\bf S},t),
\label{eq:expansion}
\end{equation}
where the deviation $\delta P$ between $P_d$ and $P_0$, which
is of order $N^0$, is of zero average: $\langle \delta P({\bf
S},t)\rangle=0$. Substituting Eq.~(\ref{eq:expansion}) in
Eq.~(\ref{eq:FPE1}), and using ${\bf h}^{\rm eff}={\bf h}^{{\rm
eff},0}[P_0]+1/\sqrt{N}~\delta {\bf h}^{\rm eff}[\delta P]$, we get 
\begin{eqnarray}
&&\fl 
\frac{\partial P_0({\bf S},t)}{\partial
t}+\frac{1}{\sqrt{N}}\frac{\partial \delta P({\bf S},t)}{\partial t} \nonumber \\
 &&\fl = -\frac{\partial}{\partial{\bf
 S}}\cdot\Big[({\bf S}\times{\bf h}^{{\rm
 eff},0})-\gamma({\bf
 S}\times{\bf S}\times{\bf h}^{{\rm eff},0})+{\cal
 D}(1+\gamma^{2})({\bf
 S}\times({\bf S}\times\frac{\partial}{\partial{\bf
 S}}))\Big]P_0\nonumber \\
 &&\fl  -\frac{1}{\sqrt{N}}\frac{\partial}{\partial{\bf
 S}}\cdot\Big[({\bf S}\times\delta{\bf h}^{{\rm
 eff}})P_0+({\bf S}\times{\bf h}^{{\rm
 eff},0})\delta P-\gamma({\bf
 S}\times{\bf S}\times\delta{\bf h}^{{\rm eff}})P_0-\gamma({\bf
 S}\times{\bf S}\times{\bf h}^{{\rm eff},0})\delta P\nonumber \\
 &&\fl+{\cal
 D}(1+\gamma^{2})({\bf
 S}\times({\bf S}\times\frac{\partial}{\partial{\bf
 S}}))\delta P\Big]-\frac{1}{N}\frac{\partial}{\partial{\bf
 S}}\cdot\Big[({\bf S}\times\delta{\bf h}^{{\rm
 eff}})\delta P-\gamma({\bf
 S}\times{\bf S}\times\delta{\bf h}^{{\rm eff}})\delta P\Big].\label{eq:FPE2}
\end{eqnarray}
Using $\langle \delta P \rangle=0$, implying $\langle \delta {\bf h}^{\rm eff}
\rangle=0$, then yields
\begin{eqnarray}
&&
\fl \frac{\partial P_0({\bf S},t)}{\partial
t} +\frac{\partial}{\partial{\bf
 S}}\cdot({\bf S}\times{\bf h}^{{\rm
 eff},0})P_0= -\frac{\partial}{\partial{\bf
 S}}\cdot\Big[-\gamma({\bf
 S}\times{\bf S}\times{\bf h}^{{\rm eff},0})+{\cal
 D}(1+\gamma^{2})({\bf
 S}\times({\bf S}\times\frac{\partial}{\partial{\bf
 S}}))\Big]P_0\nonumber \\
 &&\fl-\frac{1}{N}\Big\langle\frac{\partial}{\partial{\bf
 S}}\cdot\Big[({\bf S}\times\delta{\bf h}^{{\rm
 eff}})\delta P-\gamma({\bf
 S}\times{\bf S}\times\delta{\bf h}^{{\rm eff}})\delta
 P\Big]\Big\rangle.\label{eq:FPE3}
\end{eqnarray}
In the limit $N \to \infty$, when the last term on the right hand side
drops out, requiring that the Boltzmann-Gibbs equilibrium state $P_0({\bf S})={\cal
N}\exp(-\beta(-{\bf S}\cdot{\bf m}[P_0]+DS_{z}^{2n}))$, with
${\cal N}$ being the normalization, solves Eq.~(\ref{eq:FPE3}) in the stationary state, we must have
\begin{equation}
\fl \frac{\partial}{\partial{\bf S}}\cdot\Big(\Big[({\bf S}\times{\bf
h}^{{\rm eff},0})-\gamma({\bf S}\times{\bf S}\times{\bf h}^{{\rm
eff},0})+{\cal D}(1+\gamma^{2})({\bf S}\times({\bf
S}\times\frac{\partial}{\partial{\bf S}}))\Big]  P_0({\bf S})\Big)=0.\label{stationary-cond}
\end{equation}
Using
\begin{eqnarray}
&&\Big({\bf S}\times({\bf S}\times\frac{\partial P_0}{\partial{\bf
S}})\Big)  = \beta P_0 \Big[ {\bf
S}\Big({\bf S}\cdot{\bf m}[P_0] + {\bf S}\cdot {\bf h}^{{\rm
aniso}} \Big)-{\bf m}[P_0] - {\bf h}^{{\rm aniso}}  \big], \\
&&\gamma({\bf S}\times{\bf S}\times{\bf h}^{{\rm
eff}})P_0  =\gamma P_0 \Big[ {\bf S}({\bf
S}\cdot{\bf m}[P_0] + {\bf S}\cdot{\bf h}^{{\rm aniso}}) -{\bf m}[P_0]
- {\bf h}^{{\rm aniso}} \Big], \\
&&\frac{\partial}{\partial{\bf S}}\cdot\Big[({\bf S}\times{\bf h}^{{\rm
eff}})P_0\Big]=({\bf S}\times{\bf h}^{{\rm eff}})\cdot\frac{\partial
P_0}{\partial{\bf S}}=\beta P_0 ({\bf S}\times{\bf h}^{{\rm eff}})\cdot
{\bf h}^{{\rm eff}}=0,
\end{eqnarray}
we see that Eq.~(\ref{stationary-cond})
is satisfied provided ${\cal D}(1+\gamma^{2})\beta=\gamma$.
Consequently, Eq.~(\ref{eq:FPE3}) may be rewritten
as
\begin{eqnarray}
&& 
\fl \frac{\partial P_0({\bf S},t)}{\partial
t} +\frac{\partial}{\partial{\bf
 S}}\cdot({\bf S}\times{\bf h}^{{\rm
 eff},0})P_0 = \gamma\frac{\partial}{\partial{\bf
 S}}\cdot\Big[({\bf
 S}\times{\bf S}\times{\bf h}^{{\rm eff},0})-(1/\beta)({\bf
 S}\times({\bf S}\times\frac{\partial}{\partial{\bf
 S}}))\Big]P_0\nonumber \\
 &&\fl-\frac{1}{N}\Big\langle\frac{\partial}{\partial{\bf
 S}}\cdot\Big[({\bf S}\times\delta{\bf h}^{{\rm
 eff}})\delta P-\gamma({\bf
 S}\times{\bf S}\times\delta{\bf h}^{{\rm eff}})\delta
 P\Big]\Big\rangle.\label{eq:FPE-final}
\end{eqnarray}
In the limit $N \to \infty$, one gets Eq.~(\ref{eq:time-evolution-stochastic}) of the main text.

Let us consider the case of deterministic dynamics ($\gamma=0$). Then,
in the limit $N \to \infty$, one
gets from Eq.~(\ref{eq:FPE-final}) the Vlasov equation, Eq~(\ref{eq:Vlasov-1}), of the main
text:
\begin{eqnarray}
&& 
\frac{\partial P_0({\bf S},t)}{\partial
t} +\frac{\partial}{\partial{\bf
 S}}\cdot({\bf S}\times{\bf h}^{{\rm
 eff},0})P_0=0.
 \label{eq:app-vlasov}
\end{eqnarray}
Alternatively, Eq.~(\ref{eq:app-vlasov}) describes for finite $N$ the
time evolution for times $t \ll N$, with that for $\delta P$ obtained
from Eq.~(\ref{eq:FPE2}) as
\begin{eqnarray}
&&\frac{\partial \delta P({\bf S},t)}{\partial t} = -\frac{\partial}{\partial{\bf
 S}}\cdot\Big[({\bf S}\times\delta{\bf h}^{{\rm
 eff}})P_0+({\bf S}\times{\bf h}^{{\rm
 eff},0})\delta P\Big].\label{eq:FPE2-again}
\end{eqnarray}
The time evolution for times of order $N$ is obtained from Eq.~(\ref{eq:FPE-final}) as
\begin{eqnarray}
&& \frac{\partial P_0({\bf S},t)}{\partial
t} +\frac{\partial}{\partial{\bf
 S}}\cdot({\bf S}\times{\bf h}^{{\rm
 eff},0})P_0=-\frac{1}{N}\Big\langle\frac{\partial}{\partial{\bf
 S}}\cdot({\bf S}\times\delta{\bf h}^{{\rm
 eff}})\delta P\Big\rangle.\label{eq:FPE-final-LB}
\end{eqnarray}

\section{Appendix C: Numerical scheme for integrating
Eq.~(\ref{eq:eom-noise})}
\label{appendix3}

Here we summarize a method~\cite{Nishino:2015} to numerically integrate the
dynamics~(\ref{eq:eom-noise}) for given values of $\gamma$, $T$ and $N$. To
integrate the dynamics over a
time interval $[0 : {\cal T}]$, we first choose a time step size $\Delta
t \ll 1$, and set $t_n = n\Delta t$ as the
$n$-th time step of the dynamics, with $n = 0, 1, 2, . . . , {\cal
N}_t$, and ${\cal N}_t = {\cal T}/\Delta t$. One step of the update scheme from $t_n$ to $t_{n+1}=t_n +\Delta t$ involves the following updates of the dynamical variables
for $i = 1, 2, \ldots, N$ and $\mu,\nu,\ldots=x,y,z$:
\begin{eqnarray}
&&\fl S_i^\mu(t_{n}+\Delta t)=S_i^\mu(t_n)+F_i^\mu\left(\{{\bf S}_i(t_n+\Delta
t/2)\}\right)\Delta t + g_i^{\mu} \left(\{{\bf S}_i(t_n+\Delta
t/2)\}\right), \\
&&\fl S_i^\mu(t_n+\Delta t/2)=S_i^\mu(t_n)+F_i^\mu\left(\{{\bf
S}_i(t_n)\}\right)\Delta t/2+g_i^{\mu}(\{{\bf
S}_i\left(t_n\right)\}), \\
&&\fl F_i^\mu\left(\{{\bf S}_i(t_n)\}\right)=\epsilon_{\mu \nu
\delta}S_i^\nu(t_n)~(h_i^{\rm
eff})^\delta\left(\{{\bf
S}_i\}(t_n)\right)-\gamma \epsilon_{\mu \nu \delta}S_i^\nu(t_n) \epsilon_{\delta
\eta \zeta}S_i^\eta (t_n)~ (h_i^{\rm eff})^\zeta \left(\{{\bf
S}_i\}(t_n)\right), \nonumber \\ \\
&&\fl g_i^{\mu} \left(\{{\bf S}_i(t_n)\}\right)=\epsilon_{\mu \nu
\delta}S_i^\nu (t_n)\Delta W_i^\delta-\gamma \epsilon_{\mu \nu \delta}S_i^\nu
\epsilon_{\delta \eta \zeta}S_i^\eta \Delta W_i^\zeta,
\end{eqnarray}
where Einstein summation convention is implied. Here, $\Delta W_i^\nu$ is a Gaussian distributed random number with zero
mean and variance equal to $2\gamma T$. 

\vspace{1cm}

\end{document}